\documentclass[aip,
 amsmath,amssymb,
 reprint,%
]{revtex4-1}

\usepackage{graphicx}
\usepackage{dcolumn}
\usepackage{bm}
\usepackage{hyperref}
\usepackage{siunitx}
\usepackage{braket}
\usepackage{placeins}
\usepackage{caption}
\usepackage{xcolor}

\begin{document}

\preprint{AIP/123-QED}

\title{Chiral magnons for spin-qubit state transfer}

\author{Martijn Dols}
\email{martijn.dols@rwth-aachen.de}
\affiliation{Institute for Theoretical Solid State Physics, RWTH Aachen University, 52074 Aachen, Germany}

\author{Carlos Gonzalez-Ballestero}
\affiliation{Institute for Theoretical Physics and Vienna Center for Quantum Science and Technology, TU Wien, 1040 Vienna, Austria}

\author{Mikhail Cherkasskii}
\affiliation{Institute for Theoretical Solid State Physics, RWTH Aachen University, 52074 Aachen, Germany}

\author{Christian L. Degen}
\affiliation{Laboratory for Solid State Physics, ETH Zürich, 8093 Zurich, Switzerland}
\affiliation{Quantum Center, ETH Zürich, 8093 Zurich, Switzerland}

\author{Victor A. S. V. Bittencourt}
\email{bittencourt@fisica.unam.mx}
\affiliation{Instituto de F\'{i}sica, Universidad Nacional Aut\'{o}noma de M\'{e}xico, 04510, Mexico City, Mexico}

\author{Silvia {Viola Kusminskiy}}
\email{kusminskiy@physik.rwth-aachen.de}
\affiliation{Institute for Theoretical Solid State Physics, RWTH Aachen University, 52074 Aachen, Germany}
\affiliation{Max Planck Institute for the Science of Light, Staudtstraße 2, 91058 Erlangen, Germany}

\date{\today}

\begin{abstract}

We propose a protocol where chiral magnons mediate a state transfer between two distant spin qubits.
The protocol is implemented by varying the coupling between the spin qubits and the magnons in time, such that an arbitrary state is transferred from one qubit to the other.
The modulation of the coupling is performed such that the two-spin-qubit state is kept as a dark state of the magnon bath, bypassing the associated losses.
We show that the protocol can be realized on a hybrid system composed of two nitrogen-vacancy (NV) centers coupled to the nonreciprocal and chiral magnon modes of an yttrium iron garnet (YIG) stripe.
We propose two methods to achieve the time modulation of the NV-magnon coupling: i) the NV-magnet distance of both NV centers is varied; ii) the external magnetic field and the NV-magnet distance of one NV center are varied.
We evaluate the implementability of both methods numerically, including the constraints on the temperature, and the dephasing time and minimal lifetime of the spin qubits required for high-fidelity state transfer.
We find that using realistic experimental parameters, a state transfer between NV centers at a distance of several microns can be achieved with a fidelity $\gtrsim 0.95$.
Our findings expand the toolbox of magnonics for quantum information purposes.

\end{abstract}
\maketitle

\section{Introduction}
Magnons, the collective excitations of ordered spin systems, have shown promising properties for integration in hybrid quantum systems, with potential for enhancing various quantum information tasks~\cite{chumakAdvancesMagneticsRoadmap2022,yuanQuantumMagnonicsWhen2022,zarerameshtiCavityMagnonics2022,flebus2024MagnonicsRoadmap2024}.
The possibility of engineering their coherent interaction with other degrees of freedom, such as optical photons~\cite{osadaCavityOptomagnonicsSpinOrbit2016,liuOptomagnonicsMagneticSolids2016,violakusminskiyCoupledSpinlightDynamics2016,zhangOptomagnonicWhisperingGallery2016,haighSelectionRulesCavityenhanced2018,bittencourtOptomagnonicsDispersiveMedia2022}, microwaves~\cite{soykalStrongFieldInteractions2010,hueblHighCooperativityCoupled2013,tabuchiHybridizingFerromagneticMagnons2014,zhangStronglyCoupledMagnons2014,goryachevHighCooperativityCavityQED2014}, superconducting circuits~\cite{tabuchiCoherentCouplingFerromagnetic2015,lachance-quirionEntanglementbasedSingleshotDetection2020,wolskiDissipationBasedQuantumSensing2020,kounalakisAnalogQuantumControl2022,xuQuantumControlSingle2023,dolsMagnonmediatedQuantumGates2024a}, single-spin systems,~\cite{trifunovicLongDistanceEntanglementSpin2013,casolaProbingCondensedMatter2018,flebusEntanglingDistantSpin2019,bertelliMagneticResonanceImaging2020,neumanNanomagnonicCavitiesStrong2020,fukamiOpportunitiesLongRangeMagnonMediated2021,gonzalez-ballesteroQuantumInterfaceSpin2022,hetenyiLongdistanceCouplingSpin2022,karanikolasMagnonmediatedSpinEntanglement2022,fukamiMagnonmediatedQubitCoupling2024,bejaranoParametricMagnonTransduction2024,pengCavityMagnonPolariton2025,xueDirectionalEntanglementSpinorbit2025,dolsSteadystateEntanglementSpin2026} and phonons~\cite{weilerSpinPumpingCoherent2012,zhangCavityMagnomechanics2016,anCoherentLongrangeTransfer2020,gonzalez-ballesteroTheoryQuantumAcoustomagnonics2020,gonzalez-ballesteroQuantumAcoustomechanicsMicromagnet2020,pottsDynamicalBackactionMagnomechanics2021,schlitzMagnetizationDynamicsAffected2022,mullerChiralPhononsPhononic2024,deySensingMagnonicQuantum2025,romlingSqueezingQuantumControl2025,bruhlmannClassicalQuantumTheory2026,hwangHarmonicSubharmonicMagnon2026} positions them as a versatile and tunable collective excitation.
Moreover, magnons exhibit phenomena such as squeezing, topology, chirality and nonreciprocity, which can be exploited to enhance or harness different types of interactions~\cite{sharmaSpinCatStates2021,hetenyiLongdistanceCouplingSpin2022,dolsMagnonmediatedQuantumGates2024a,xueDirectionalEntanglementSpinorbit2025,romlingSqueezingQuantumControl2025,mycroftQuantumStatePreparation2026,dolsSteadystateEntanglementSpin2026}.

A central ingredient in quantum information processing is the deterministic transfer of an unknown qubit state between two spatially separated nodes.
In one-dimensional quantum channels this can be achieved when (i) the excitations of the channel propagate unidirectionally from the sender to the receiver and (ii) the node–channel coupling is tunable in time, so that the emitted wave packet is shaped to be perfectly reabsorbed~\cite{ciracQuantumStateTransfer1997}.
Such ideas are well established in chiral quantum optics~\cite{lodahlChiralQuantumOptics2017,suarez-foreroChiralQuantumOptics2025}, where unidirectional light-matter coupling enables state-transfer protocols based on engineered, time-dependent decay rates~\cite{ciracQuantumStateTransfer1997,stannigelOptomechanicalTransducersQuantuminformation2011a,vermerschQuantumStateTransfer2017,xiangIntracityQuantumCommunication2017}.
However, combining strong directionality with broad in-situ tunability is often challenging in purely photonic implementations~\cite{awschalomQuantumTechnologiesOptically2018a}.
This motivates using magnons as directional mediators, since their chirality and nonreciprocity can emerge intrinsically on various magnetic systems, including 
ferromagnetic slabs~\cite{damonMagnetostaticModesFerromagnet1961,parekhPropagationCharacteristicsMagnetostatic1985}, ultra-thin layers~\cite{udvardiChiralAsymmetrySpinWave2009,cortes-ortunoInfluenceDzyaloshinskiiMoriya2013,belmeguenaiInterfacialDzyaloshinskiiMoriyaInteraction2015,tacchiInterfacialDzyaloshinskiiMoriyaInteraction2017,gallardoFlatBandsIndirect2019}, and nanotubes~\cite{otaloraCurvatureInducedAsymmetricSpinWave2016,korberCurvilinearSpinwaveDynamics2022}, as well as ferromagnetic~\cite{hillebrandsSpinwaveCalculationsMultilayered1990,zakeriAsymmetricSpinWaveDispersion2010,gallardoSpinwaveNonreciprocityMagnetizationgraded2019,gallardoReconfigurableSpinWaveNonreciprocity2019,gallardoCoherentMagnonsGiant2024,heinsNonreciprocalSpinwaveDispersion2025} and antiferromagnetic~\cite{wintzMagneticVortexCores2016,slukaEmissionPropagation1D2019,ishibashiSwitchableGiantNonreciprocal2020,thiancourtUnidirectionalSpinWaves2024,wojewodaUnidirectionalPropagationZeromomentum2024} multilayered structures.
The in-situ tunability of key parameters such as dispersion and coupling strengths~\cite{yuanQuantumMagnonicsWhen2022,zarerameshtiCavityMagnonics2022} provides a flexible route to realize—and potentially extend—unidirectional state-transfer protocols.
Moreover, the relatively short magnon wavelength, typically in the micron regime~\cite{chumakMagnonSpintronics2015}, can help to circumvent bulky hybrid systems, possibly paving the way towards on-chip architectures.

In this work, we show that the nonreciprocity and chirality of magnons combined with their tunability can be used to perform a high-fidelity state transfer between distant spin qubits.
The nonreciprocity and chirality of magnons gives rise to an effective unidirectional coupling between spin qubits~\cite{xueDirectionalEntanglementSpinorbit2025,dolsMagnonmediatedQuantumGates2024a}. 
We show that this effective qubit-qubit coupling can be modulated thanks to the tunability of magnons, in a way that enables state transfer between two distant qubits.
The protocol can be performed such that the two-qubit state remains a dark state of the effective two-qubit dynamics, bypassing dissipation into the magnon bath.
We numerically test the state-transfer protocol on a hybrid quantum system where nitrogen-vacancy (NV) centers~\cite{weberQuantumComputingDefects2010,dohertyNitrogenvacancyColourCentre2013} are coupled to the nonreciprocal and chiral magnon modes of a yttrium iron garnet (YIG) stripe~\cite{casolaProbingCondensedMatter2018,bertelliMagneticResonanceImaging2020,fukamiMagnonmediatedQubitCoupling2024}.
In such a hybrid system, the qubit-magnon coupling can be modulated by controlling the distance between the NVs and the stripe, or by varying the external bias field, which changes the magnon dispersion.
We show how both effects can be exploited to implement the protocol which, under realistic experimental conditions, can achieve a state transfer fidelity of $\gtrsim0.95$.
It is required that the spin qubits are placed within the coherence length of the magnons, reaching hundreds of microns in state-of-the-art setups~\cite{bensmannDispersiontunableLowlossImplanted2025,serhaUltralonglivingMagnonsQuantum2025}.

The remainder of this work is structured as follows.
In Sec.~\ref{sec:model} we present the master equation of two spin qubits coupled unidirectionally through a nonreciprocal and chiral magnon bath.
The state-transfer protocol is discussed and benchmarked with respect to the temperature and the intrinsic qubit dissipation in Sec~\ref{sec:prot}.
We discuss in Sec.~\ref{sec:imp} the implementation of the protocol on the NV-YIG hybrid system: in Sec.~\ref{sec:d_1_2} we show our results of the implementation by modulating the distance of both NV centers to the magnetic strip, whereas the variation of the magnetic field and of one NV-YIG distance is presented in Sec.~\ref{sec:H_d}.
We present the conclusions of this work in Sec.~\ref{sec:con}.

\begin{figure}[ht]
\includegraphics[width=\linewidth]{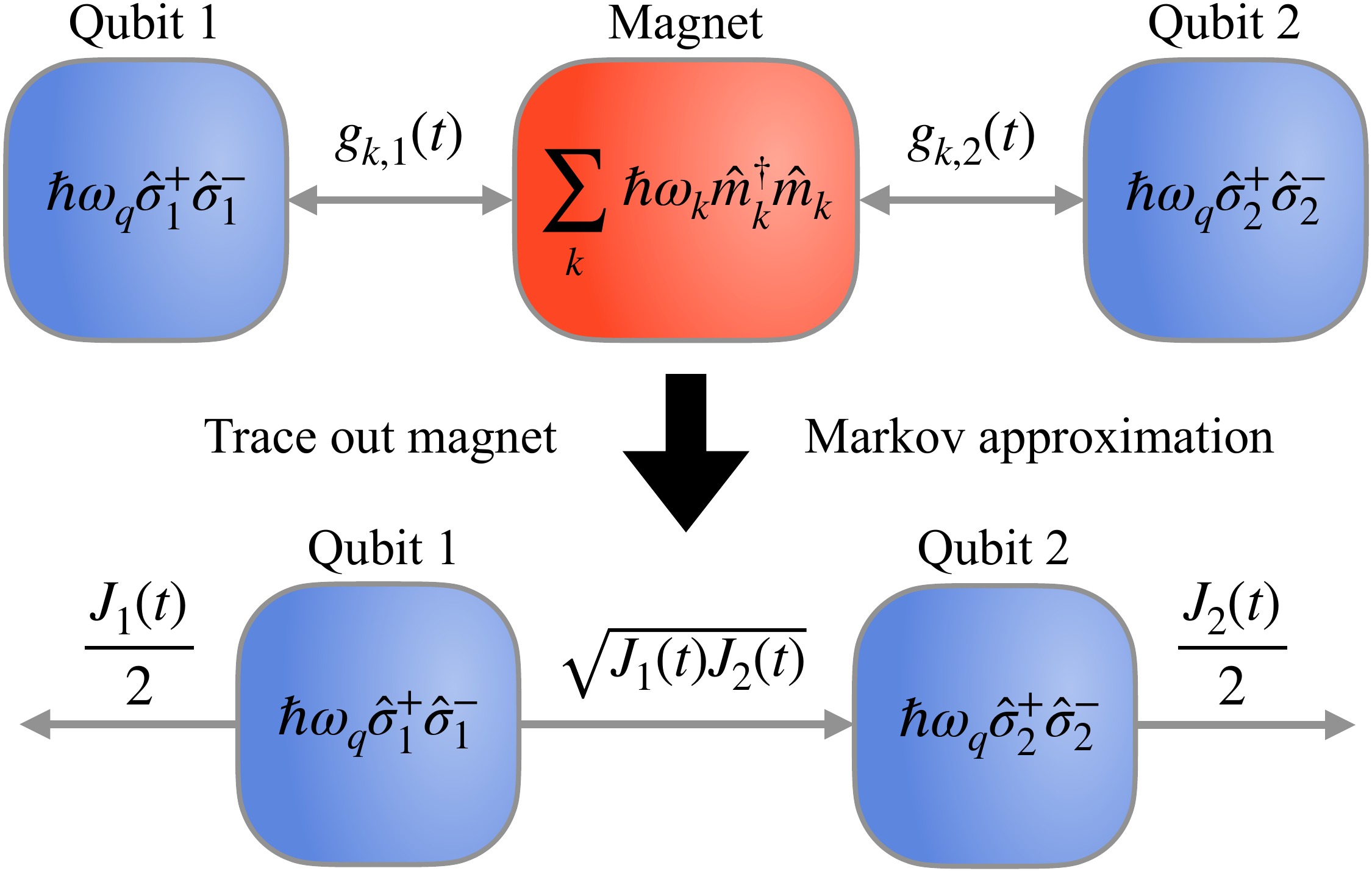}
\caption{\label{fig:coupling} Two qubits $i=1,2$ are coupled to a magnon bath with the time-dependent coupling $g_{k,i}(t)$ limited to one propagation direction of the magnons.
The effective qubit-qubit coupling is unidirectional, such that the dynamics of qubit 2 are influenced by qubit 1, but not vice versa.
The qubits dissipate into the bath at a rate $J_{i}(t)/2$.}
\end{figure}

\section{Model}\label{sec:model}
We consider a system of two spin qubits coupled to a magnon bath as schematically shown in Fig.~\eqref{fig:coupling}. The two identical qubits $i=1,2$ with a transition frequency $ \omega_q$ between the computational states $\ket{0}_{i}$ and $\ket{1}_{i}$ are modeled by the Hamiltonian
\begin{equation}
    \hat{H}_{Q}=\hbar\omega_{q}(\hat{\sigma}_{1}^{+}\hat{\sigma}_{1}^{-}+\hat{\sigma}_{2}^{+}\hat{\sigma}_{2}^{-}),
\label{eq:H_Q}
\end{equation}
where $\hat{\sigma}_{i}^{-} = \ket{0}_{i}\bra{1}_{i}$ and $\hat{\sigma}_{i}^{+} = (\hat{\sigma}_{i}^{-})^{\dagger}$.

Magnons are the bosonic quasiparticles that describe low-energy excitations in magnetic systems. Within linear spin-wave theory, the magnon Hamiltonian takes the form~\cite{stancilSpinWavesTheory2009}
\begin{equation}
    \hat{H}_{M}=\sum_{k}\hbar\omega_{k}\hat{m}_{k}^{\dagger}\hat{m}_{k},
\label{eq:H_M}
\end{equation}
where $\hat{m}_{k}$ is the annihilation operator of a magnon mode $k$.
The magnon modes and their dispersion relation are determined by the intrinsic properties of the magnetic material (saturation magnetization, g-factor, etc.), the geometry of the magnet, and the applied magnetic field.

The magnetic moments of the spin qubits couple to the fluctuating magnetic fields produced by magnons, resulting in a dipolar interaction. Within the rotating-wave approximation (RWA), the interaction Hamiltonian reads~\cite{trifunovicLongDistanceEntanglementSpin2013,flebusEntanglingDistantSpin2019,bertelliMagneticResonanceImaging2020,neumanNanomagnonicCavitiesStrong2020,fukamiOpportunitiesLongRangeMagnonMediated2021,gonzalez-ballesteroQuantumInterfaceSpin2022,hetenyiLongdistanceCouplingSpin2022,karanikolasMagnonmediatedSpinEntanglement2022,fukamiMagnonmediatedQubitCoupling2024,bejaranoParametricMagnonTransduction2024,xueDirectionalEntanglementSpinorbit2025}
\begin{equation}
\hat{H}_{\mathrm{int}}(t)=\sum_{k}\hbar\left(g_{k,1}(t)\hat{\sigma}_{1}^{+}+g_{k,2}(t)\hat{\sigma}_{2}^{+}\right)\hat{m}_{k}+\mathrm{H.c.}
\label{eq:H_int}
\end{equation}
The unidirectionality of the coupling of the spin qubits to the magnons is encoded in the coupling strength $g_{k,i}(t)$, which we assume to be time-dependent (see Sec.~\ref{sec:imp}).
In the case of chiral magnons, the dipolar interaction between the chiral stray field of the magnons with the magnetic moment of the spin gives rise to a selection rule associated to the propagation direction of the magnons.
Nonreciprocal magnons either merely exist in one propagation direction, or their frequency dispersion is asymmetric, such that only one magnon propagation direction couples resonantly to the spin qubit.
Such phenomena can be used to generate a (uni)directional coupling between spin qubits mediated by magnons~\cite{gonzalez-ballesteroQuantumInterfaceSpin2022,xueDirectionalEntanglementSpinorbit2025,dolsSteadystateEntanglementSpin2026}.
We will show a specific example of qubits coupling to chiral and nonreciprocal magnons in Sec.~\ref{sec:imp}.

The qubits $1$ and $2$ are positioned at $r_1$ and $r_2$, such that the distance between them is $r_{2,1}=r_2 - r_1$.
The coupling to the magnons induces an effective qubit-qubit interaction, which can be obtained by tracing over the magnon modes within the Markov approximation.
To this end, we follow closely our derivation in Ref.~\onlinecite{dolsSteadystateEntanglementSpin2026}, taking additionally into account that in the present work the magnon-qubit coupling is time-dependent.
We present the detailed derivation in App.~\ref{app:EOM}.
In the low-temperature regime, characterized by $k_B \mathcal{T} \ll \hbar \omega_q$ with temperature $\mathcal{T}$, the master equation describing two spin qubits coupled unidirectionally through the magnon bath is given by~\cite{dolsSteadystateEntanglementSpin2026}
\begin{equation}
\begin{aligned}
    \frac{\mathrm{d}\hat{\rho}}{\mathrm{d} t} & =-\frac{i}{\hbar}\left[\hat{H}_{Q},\hat{\rho}\right]-\frac{i}{\hbar}\left(\hat{H}_{\mathrm{eff}}(t)\hat{\rho}-\hat{\rho}\hat{H}_{\mathrm{eff}}^{\dagger}(t)\right) \\ & \quad +\hat{L}(t)\hat{\rho}\hat{L}^{\dagger}(t),
\end{aligned}
\label{eq:EOM}
\end{equation}
where $\hat{H}_{\mathrm{eff}}(t) = \hat{H}_{\mathrm{loc}}(t)+ \hat{H}_{\mathrm{uni}}(t)$ consists of a local, non-Hermitian Hamiltonian describing the dissipation of the qubits into the bath, 
\begin{equation}
    \hat{H}_{\mathrm{loc}}(t)=-\frac{i\hbar}{2}\left(J_{1}(t)\hat{\sigma}_{1}^{+}\hat{\sigma}_{1}^{-}+J_{2}(t)\hat{\sigma}_{2}^{+}\hat{\sigma}_{2}^{-}\right),
\label{eq:H_loc}
\end{equation}
and a Hamiltonian yielding a unidirectional coupling between qubit 1 and 2, 
\begin{equation}
    \hat{H}_{\mathrm{uni}}(t)=-i\hbar\sqrt{J_{1}(t)J_{2}(t)}e^{ik_{q}r_{2,1}}\hat{\sigma}_{1}^{-}\hat{\sigma}_{2}^{+}.
\label{eq:H_uni}
\end{equation}
Here, we defined the dissipative coupling constant $J_{i}(t)=|g_{i,k_{q}}(t)|^{2} L/v_{k_q}$ and the wave number $k_q$ which is such that the magnons and the qubits are resonant ($\omega_{k_q} = \omega_q$), with $v_k = \partial\omega_k/\partial k$ being the group velocity and $L$ denoting the quantization length of the magnetic stripe along the magnon-propagation direction, i.e. the direction along which the two NV centers are separated.
The jump operator $\hat{L}(t)$ is given by
\begin{equation}
    \hat{L}(t) = \sqrt{J_{1}(t)}e^{-ik_{q}r_{1}}\hat{\sigma}_{1}^{-} + \sqrt{J_{2}(t)}e^{-ik_{q}r_{2}}\hat{\sigma}_{2}^{-}.
\label{eq:L}
\end{equation}
The validity of the master equation~(\ref{eq:EOM}) is given by the conditions of the Markov approximation, which rely on the short correlation time of the magnon bath with respect to the qubits~\cite{breuerTheoryOpenQuantum2007}.
Accordingly, it is assumed that $\tau_m \ll \tau_{q},\,\tau_g$, with $\tau_m$ and $\tau_q $ being the correlation time of the magnon bath and of the spin qubit, respectively, while $\tau_g$ is a time constant associated with the rate at which the NV-magnon coupling $ g_{{k},i}(t)$ changes in time, i.e. $\tau_g \sim |g_{{k},i}(t)/\dot{g}_{{k},i}(t)|$.
The condition $\tau_m \ll \tau_g$ ensures that the variation of $ g_{{k},i}$ in time does not introduce any non-Markovianity due to a significant variation within the bath's correlation time.
Furthermore, it is assumed that there is no time retardation between the qubits which are coupled through the bath, i.e. $ J_{1}(t),J_{2}(t) \ll v_{k_q} / r_{2,1}$, where $r_{2,1} / v_{k_q}$ is the time at which the quanta emitted by the left qubit arrive at the right qubit through the bath. 
In particular, if $r_{2,1} / v_{k_q}$ would be comparable with $1/J_{1,2}(t)$,
retardation effects can no longer be neglected and the dynamics deviate from a time-local (Markovian) description~\cite{Gonzalez-Ballestero_2013}.
Finally, the distance $r_{2,1}$ between the qubits may not exceed the coherence length $l_m = v_{k_q} \tau_m$ of the magnons.

Typically, the jump term $\hat{L} \hat{\rho} \hat{L}^{\dagger}$ of the master equation Eq.~(\ref{eq:L}) causes loss of excitations and decoherence of the qubits due to the interaction with the bath.
In the next section~\ref{sec:prot}, we use the time dependence of the dissipative couplings $J_{1,2}(t)$ to let the two-qubit state evolve in the null space of the jump operator, i.e. the dark-state condition, to prevent such effects.

\section{Protocol}\label{sec:prot}
In this section, we show how the dissipative couplings $J_{1,2}(t)$ should be tuned such that a state transfer between qubit $1$ and $2$ is performed, while the two-qubit state remains a dark state~\cite{ciracQuantumStateTransfer1997,stannigelOptomechanicalTransducersQuantuminformation2011a}.
We consider an initial state at time $t_{i}$ given by $\ket{\psi(t_{i})}=\ket{\psi}_{1}\ket{0}_{2}$, where $\ket{\psi}_{1}=\alpha\ket{0}_{1}+\beta\ket{1}_{1}$.
To achieve the state transfer, the evolved state at the final time $t_{f}$ has to be $\ket{\psi_{T}}=\ket{0}_{1}\ket{\psi}_{2}$.
For this sake, we take the ansatz
\begin{equation}
\begin{aligned}
    \ket{\psi(t)}&=\alpha c_{00}(t)\ket{0}_{1}\ket{0}_{2}\\ &\quad +\beta e^{-i\varphi(t)}\left(c_{10}(t)\ket{1}_{1}\ket{0}_{2}+c_{01}(t)\ket{0}_{1}\ket{1}_{2}\right),
\end{aligned}
\label{eq:psi_t}
\end{equation}
with $\varphi(t)=\omega_{q}(t-t_i)$ and where initially $c_{00}(t_{i})=c_{10}(t_{i})=1$
and $c_{01}(t_{i})=0$.
The goal is to shape the time-dependence of the couplings $J_{1,2}(t)$, such that at the final time $c_{00}(t_{f})=c_{01}(t_{f})=1$
and $c_{10}(t_{f})=0$.
Since the initial state has no occupation in the $\ket{1}_{1}\ket{1}_{2}$ state and this state is decoupled from the dynamics, we disregard the occupation $c_{11}(t)$.
The normalization condition for $\ket{\psi(t)}$ leads to
\begin{equation}
    |\alpha c_{00}(t)|^{2}+|\beta|^{2}\left(|c_{10}(t)|^{2}+|c_{01}(t)|^{2}\right)=1.
\end{equation}
The dark-state condition $\hat{L}(t)\ket{\psi(t)}=0$ gives rise to
\begin{equation}
    \sqrt{J_{1}(t)}c_{10}(t)e^{-ik_{q}r_{1}}+\sqrt{J_{2}(t)}e^{-ik_{q}r_{2}}c_{01}(t)=0.
\label{eq:dark}
\end{equation}
Imposing that the dark-state condition is fulfilled at all times
$t$, the state $\ket{\psi(t)}$ evolves according to the Schrödinger equation $i\hbar\frac{\mathrm{d}}{\mathrm{d}t}\ket{\psi(t)}=(\hat{H}_{Q}+\hat{H}_{\mathrm{eff}}(t))\ket{\psi(t)}$.
Using the Hamiltonians of Eqs.~(\ref{eq:H_Q}),~(\ref{eq:H_loc}) and~(\ref{eq:H_uni}) together with the state given in Eq.~(\ref{eq:psi_t}) leads to
\begin{align}
    \dot{c}_{00}(t) & =0, \\
    \dot{c}_{10}(t) & =-\frac{J_{1}(t)}{2}c_{10}(t), \label{eq:c_10} \\
    \dot{c}_{01}(t) & =-\frac{J_{2}(t)}{2}c_{01}(t)-\sqrt{J_{1}(t)J_{2}(t)}e^{ik_{q}r_{2,1}}c_{10}(t). \label{eq:c_01}
\end{align}
A comparison of Eq.~(\ref{eq:c_10}) with~(\ref{eq:c_01}) demonstrates the unidirectionality of the coupling, since the occupation of the second qubit ${c}_{01}(t)$ can be influenced by the occupation of the first qubit $c_{10}(t)$, but not vice versa.

To perform the state transfer, we assume a function $f(t)$ which obeys 
\begin{equation}
c_{10}(t)=\sqrt{1-f(t)},\quad c_{01}(t)=-e^{ik_{q}r_{2,1}}\sqrt{f(t)}.
\label{eq:c_f}
\end{equation}
The function $f(t)$ is chosen such that the boundary conditions at $t=t_i$ and $t=t_f$ of $c_{00}(t)$, $c_{10}(t)$ and $c_{01}(t)$ are fulfilled, which corresponds to $f(t_{i})=0$ and $f(t_{f})=1$,
and such that the dark-state condition is satisfied at all times $t$.
Furthermore, to satisfy $c_{01}(t_{f})=1$ the phase $k_{q}r_{2,1}$ is chosen such that $k_{q}r_{2,1}=\pi+2\pi n$ with integer $n$.
Note that $k_q r_{2,1}$ can also be used to induce a local phase between the components of the final state of the second qubit, e.g. $\alpha \ket{0}_{2} - \beta \ket{1}_{2}$.
In App.~\ref{app:pulses} we find that the couplings $J_{1,2}(t)$ which comply with these constraints are given by
\begin{equation}
    J_{1}(t)=\frac{\dot{f}(t)}{1-f(t)},\quad J_{2}(t) =\frac{\dot{f}(t)}{f(t)}.
\label{eq:J_1_2}
\end{equation}

\begin{figure*}[ht]
\includegraphics[width=\textwidth]{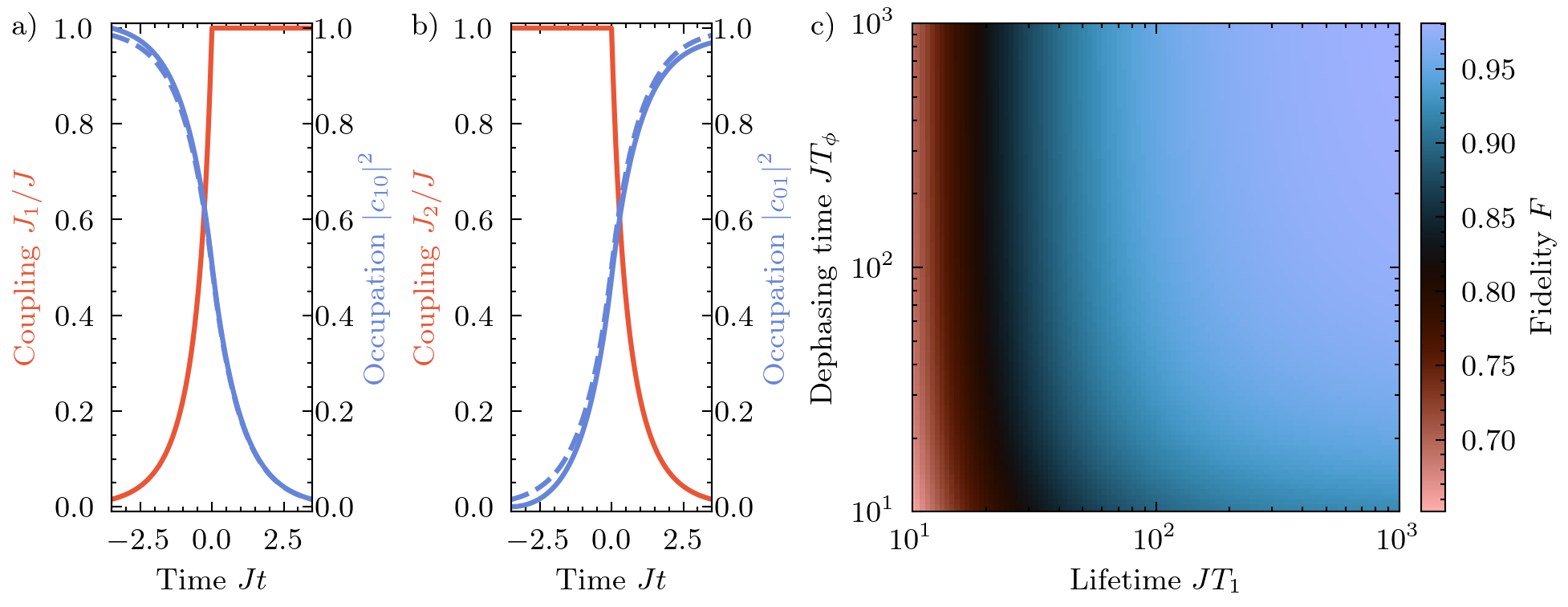}
\caption{\label{fig:protocol} In red: effective coupling a) $J_1$ and b) $J_2$ normalized by its maximal value $J$ as a function of time $t$ normalized by $J$.
In blue: simulated occupation (solid line) a) $|c_{10}(t)|^{2}$ and b) $|c_{01}(t)|^{2}$ and the ideal occupation as per Eq.~(\ref{eq:c_f}) (dashed line) a) $1-f(t)$ and b) $f(t)$.
c) Fidelity of the final state using the effective couplings of Eq.~(\ref{eq:J_1_2}) as a function of the lifetime $T_1$ and dephasing time $T_\phi$ of the qubit normalized by the maximal coupling $J$.
In all figures we used $J t_f = 3.5$, $\alpha=0$ and $\beta = 1$.}
\end{figure*}

Having obtained the effective couplings as a function of $f(t)$, one needs to determine the protocol time $t_p$ during which the couplings will be varied.
We use $t_{f}=t_p /2 =-t_{i}>0$ and set $t_{f}$ such that the fidelity of the state $\ket{\psi(t_{f})}$ with the target state $\ket{\psi_{T}}$ reaches a specific threshold $F_{T}=|\braket{\psi_{T}|\psi(t_{f})}|$. This yields the following relation\footnote{Here, we disregard the phase $\varphi(t)$ which comes from the free evolution of the qubit according to $\hat{H}_{Q}$.}
\begin{equation}
    F_{T} = |\alpha|^{2}+|\beta|^{2}\sqrt{f(t_{f})},
\label{eq:F_t_f}
\end{equation}
which can be inverted to obtain $t_f$.
A function $f(t)$ that varies from $0$ to $1$ quickly as a function of $t$ is preferred to minimize the protocol time $t_p$, while complying with the constraints on the time modulation of the coupling as discussed in Sec.~\ref{sec:model}.

If one adopts
\begin{equation}
    f(t)=
\begin{cases}
\frac{1}{2}e^{Jt}, & t<0\\
1-\frac{1}{2}e^{-Jt}, & t\geq0,
\end{cases}
\label{eq:f}
\end{equation}
the effective couplings take following simple form~\cite{stannigelOptomechanicalTransducersQuantuminformation2011a}
\begin{equation}
    J_{1}(t)=\begin{cases}
\frac{Je^{Jt}}{2-e^{Jt}}, & t<0\\
J, & t\geq0,
\end{cases}
\quad
J_{2}(t) = 
\begin{cases}
J, & t<0\\
\frac{Je^{-Jt}}{2-e^{-Jt}}, & t\geq0.
\end{cases}
\label{eq:J_t}
\end{equation}
As one can see in Fig.~\ref{fig:protocol}a), the coupling $J_1(t)$ rises for $t<0$ to a maximum $J$, at which it remains for $t\geq0$.
The coupling $J_2(t)$ takes a time-inverted form, as displayed in Fig.~\ref{fig:protocol}b).
For this coupling, the occupations $|c_{10}(t)|^2$ and $|c_{01}(t)|^2$ (solid lines in Fig.~\ref{fig:protocol}) are expected to follow $1-f(t)$ and $f(t)$ (dashed lines in Fig.~\ref{fig:protocol}), respectively. We obtain the occupations $|c_{10}(t)|^2$ and $|c_{01}(t)|^2$ by solving the master equation~(\ref{eq:EOM}) using the input state $\ket{1}_{1}\ket{0}_{2}$ and the couplings given in Eq.~(\ref{eq:J_t}).
The small deviation between the occupations and their ideal forms dictated by $f(t)$ is due to the finite protocol time $t_p$, since at the initial time $t_i = -3.5/J$, $f(t_i) \approx 0.01 $, while the boundary conditions equal $c_{10}(t_i)=0$ and $c_{01}(t_i)=1$ with $f(t_i)=0$.

According to Eq.~(\ref{eq:F_t_f}), we find that the protocol time to reach the fidelity $F_T$ with $f(t)$ given in Eq.~(\ref{eq:f}) equals
\begin{equation}
    t_{p}=-\frac{2}{J}\ln\left[2\left(1-\frac{\left(F_{T}-|\alpha|^{2}\right)^{2}}{|\beta|^{4}}\right)\right].
\end{equation}
For $\alpha=0$, $\beta = 1$ and $F_T = 0.99$, one finds $t_{p}\approx 6/J$.

The couplings $J_{1,2}(t)$ as given in Eq.~\eqref{eq:J_t}  start or finish at a finite value.
Depending on the implementation, it might be advantageous to have the qubits initially decoupled.
Thus, we look for couplings which vanish initially at $t=t_i$ and at $t=t_f$.
Setting $f(t) = 1 / (1+e^{-g(t)})$ gives rise to
\begin{equation}
    J_{1}(t)=f(t)\dot{g}(t),\quad J_{2}(t)=\left(1-f(t)\right)\dot{g}(t).
\label{eq:J_1_2_g}
\end{equation}
Effective couplings with the right boundary conditions can then be obtained by choosing a smooth function $g(t)$ which obeys $\dot{g}(t_{i})=\dot{g}(t_{f})=0$.
We set
\begin{equation}
    g(t)=A\left(\frac{t}{t_{f}}+\frac{1}{\pi}\sin\left(\frac{\pi t}{t_{f}}\right)\right),
\label{eq:g_t}
\end{equation}
where $A$ is a constant which is chosen such that a certain fidelity threshold $F_T$ is reached at $t=t_f$. With Eq.~(\ref{eq:F_t_f}) one finds for $\alpha=0$ and $\beta=1$ that $A=\ln\left(F_{T}^{2}/(1-F_{T}^{2})\right)$.
In App.~\ref{app:pulse_g} we show that the duration of the protocol is $t_p \approx 12/J$ for $F_T = 0.99$, where $J$ is the maximal value of the coupling $J_{1,2}(t)$.
Thus, having the effective coupling with vanishing initial and final values comes at the cost of a longer protocol time.

During the protocol, dissipative processes such as intrinsic qubit dephasing and decay influence the time dynamics.
Therefore, the equation of motion of Eq.~(\ref{eq:EOM}) is extended by two dissipators $1/T_{\phi} \sum_i\mathcal{D}[\hat{\sigma}_{i}^{+}\hat{\sigma}_{i}^{-}]\hat{\rho}$ and $1/T_{1} \sum_i\mathcal{D}[\hat{\sigma}_{i}^{-}]\hat{\rho}$ to model qubit dephasing and decay, respectively, where
\begin{equation}
    \mathcal{D}[\hat{A}]\hat{\rho} = \hat{A} \hat{\rho}\hat{A}^{\dagger} - \frac{1}{2}\left(\hat{A}^{\dagger}\hat{A} \hat{\rho} + \hat{\rho}\hat{A}^{\dagger}\hat{A} \right).
\end{equation}
To benchmark the dissipation with respect to the coupling $J$, we take the input state $\ket{\psi(t_i)}=\ket{1}_{1}\ket{0}_{2}$ and propagate this state until the time $t=t_f$ according to the master equation~(\ref{eq:EOM}) including the dissipators describing qubit decay and dephasing, while using the effective couplings given in Eq.~(\ref{eq:J_t}).
This gives rise to the final state $\hat{\rho}(t_f)$.
We vary the dephasing time $T_{\phi}$ and the lifetime $T_{1}$ and compute the fidelity $F=\sqrt{ \bra{\psi_T}\hat{\rho}(t_f)\ket{\psi_T} }$ of the state $\hat{\rho}(t_f)$ with the target state $\ket{\psi_T} = \ket{0}_{1}\ket{1}_{2}$.
This results in Fig.~\ref{fig:protocol}c).

\begin{figure}[h]
\includegraphics[width=\linewidth]{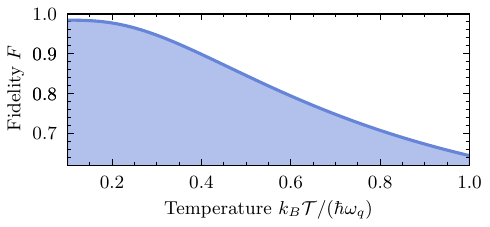}
\caption{\label{fig:temp} Fidelity $F$ of the protocol as a function of the temperature $\mathcal{T}$ normalized by $\hbar \omega_q / k_B$.
The same parameters as in Fig.~\ref{fig:protocol} are used.}
\end{figure}

The master equation shown in Sec.~\ref{sec:model} is valid for temperatures $k_B \mathcal{T} \ll \hbar \omega_q$ as shown in App.~\ref{app:EOM}.
As the temperature increases, thermal excitation of the qubits due to the finite temperature of the bath.
This, in turn, degrades the fidelity of the protocol, as shown in Fig~\ref{fig:temp}.
Here, we use the finite-temperature model (see App.~\ref{app:EOM}) instead of the zero-temperature model (see Sec.~\ref{sec:model}).
For temperatures $k_B\mathcal{T}\lesssim 0.2 \hbar \omega_q$ the thermal excitations have a negligible effect on the fidelity of the protocol.

\section{Implementation}\label{sec:imp}

\begin{figure}[ht]
\includegraphics[width=\linewidth]{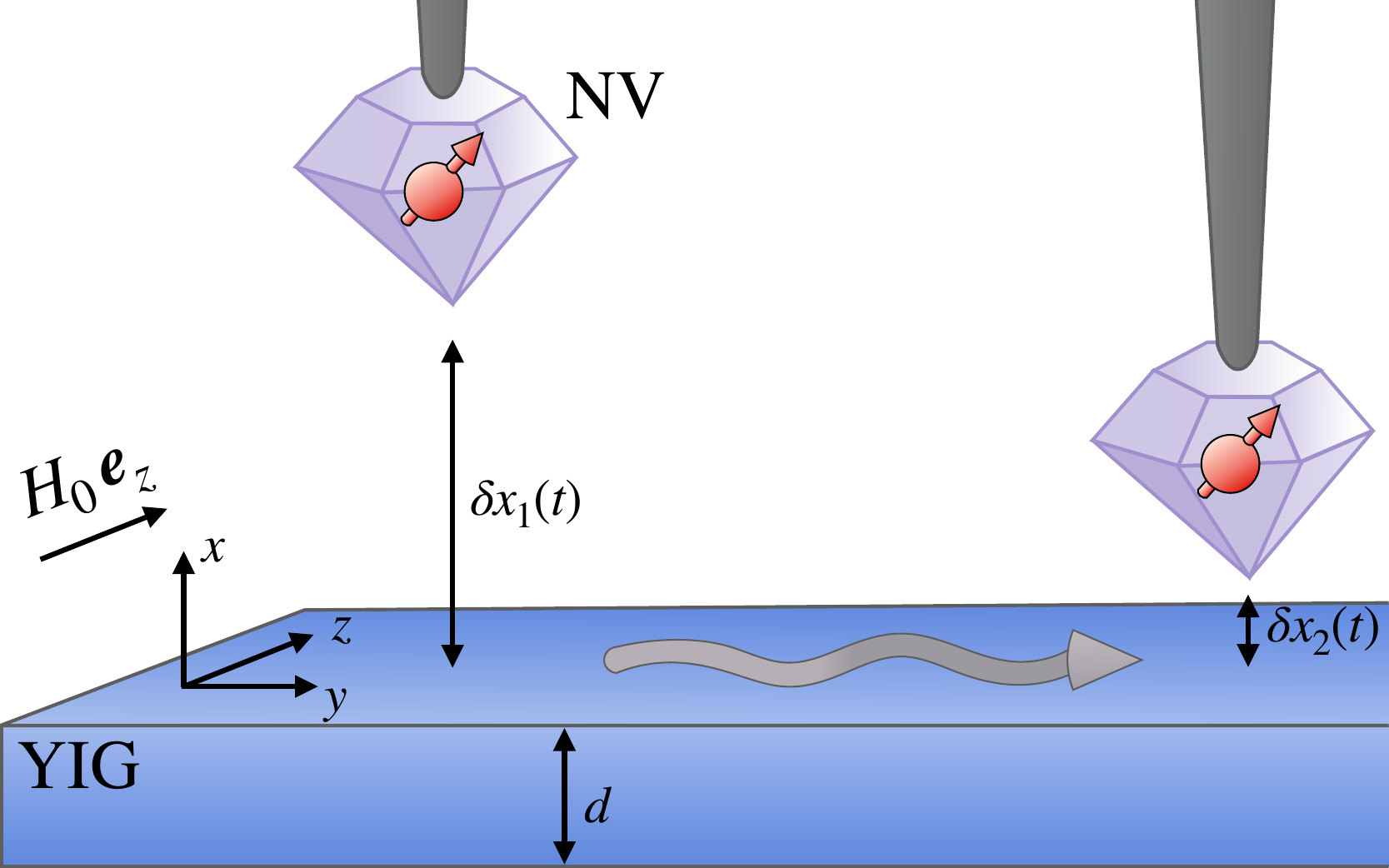}
\caption{\label{fig:setup} The proposed setup. Two NV centers integrated with a tip are coupled unidirectionally through the chiral and nonreciprocal magnon surface modes of a YIG stripe with thickness $d$.
The coupling of the NV center $i$ to the magnons can be varied through the external field $H_0(t)$ as well as the position $\delta x_i (t)$. The magnons propagate perpendicular to the external field $H_0 \boldsymbol{e}_z$.}
\end{figure}

In this section, we numerically test the protocol on a hybrid system consisting of two NV centers coupled to the surface modes of a ferromagnetic thin YIG stripe~\cite{casolaProbingCondensedMatter2018,bertelliMagneticResonanceImaging2020,fukamiMagnonmediatedQubitCoupling2024}.
NV centers are a promising platform for quantum technologies due to their comparatively long coherence times~\cite{bar-gillSolidstateElectronicSpin2013,yamamotoTenSecondElectronSpinCoherence2026} and their compatibility with hybrid architectures involving, e.g., photons~\cite{gonzalez-tudelaLightMatterInteractions2024} and phonons~\cite{yinHybridOptomechanicalSystems2015}.
YIG is the magnetic material of choice in current state-of-the-art experiments, due to its record-low dissipation values for magnons~\cite{cherepanovSagaYIGSpectra1993,serhaUltralonglivingMagnonsQuantum2025}.
The surface modes of the magnet show both chiral as well as nonreciprocal features~\cite{damonMagnetostaticModesFerromagnet1961,parekhPropagationCharacteristicsMagnetostatic1985}, which can be used to generate a unidirectional coupling between the NV centers~\cite{xueDirectionalEntanglementSpinorbit2025,dolsSteadystateEntanglementSpin2026}.
Fig.~\ref{fig:setup} shows the configuration of this hybrid system. The distance of the NV centers to the magnetic surface as well as the external field $H_0$ can be varied to tune the coupling of the NV centers to the magnons.

The ground-state manifold of an NV center (spin $S = 1$) comprises the spin-triplet states defined by the magnetic quantum numbers $m_s = 0,\pm1$, corresponding to the states $\ket{0}$ and $\ket{\pm}$. 
At a vanishing magnetic field, the energy of the $\ket{\pm}$ states is $\hbar D_0$ higher than the ground state $\ket{0}$.
An external magnetic field $H_0$ lifts the degeneracy of the $\ket{\pm}$ states by a Zeeman shift $\pm \omega_H$, with $\omega_{H}=\gamma_{s}\mu_{0}H_{0}$, where $\gamma_{s}$ is the gyromagnetic ratio and $\mu_{0}$ is the vacuum magnetic permeability.
Thus, the Hamiltonian of two identical NV centers is given by
\begin{equation}
    \hat{H}_{\mathrm{NV}} = \sum_{i=1,2} \hbar (\omega_{+}\hat{\sigma}^{++}_{i} + \omega_{-}\hat{\sigma}^{--}_{i}) ,
\end{equation}
where $\omega_{\pm}=D_{0}\pm\omega_{H}$ and $\hat{\sigma}^{\alpha \beta}_{i}=\ket{\alpha}_{i}\bra{\beta}_{i}$.

We consider a YIG stripe with dimensions $L_z \gg L_y \gg d$.
We neglect the higher-order magnon modes with $k_{z}>0$ given that they are far detuned from the NV frequencies, $g_{n_{z}=1,k} \ll |\omega_{n_{z}=1,k}-\omega_{q}|$, where $n_{z}$ is defined by $k_z=2\pi n_z/L_z$ and $k$ is the wave number $\boldsymbol{k}=k\boldsymbol{e}_y$.
This enables the limitation to surface magnons propagating along the $y$ axis with $k_{z}=0$.
The magnon surface modes which propagate along the $y$ axis perpendicular to the external field $H_0 \boldsymbol{e}_{z}$ are known as the Damon-Eshbach (DE) modes~\cite{damonMagnetostaticModesFerromagnet1961}.
The modal profiles of the DE modes are chiral, where the mode profiles above the YIG stripe obey $\delta \boldsymbol{H}_k \propto \boldsymbol{e}_{\mp}$ for $k \gtrless 0$, see also App.~\ref{app:magnon}.
Moreover, the field displacement nonreciprocity favors right-propagating magnons ($k>0$) over left-propagating ($k<0$). 

The two NV centers are positioned above the magnetic stripe along the $y$ axis at positions $\boldsymbol{r}_{i}$. As explained in Sec.~\ref{sec:model}, the distance between the spin qubits cannot exceed the coherence length $l_m = v_{k_{q}} \tau _m $ of the magnons.
Using $\tau_m = \SI{1}{\micro s}$~\cite{serhaUltralonglivingMagnonsQuantum2025} gives rise to $l_m = \SI{0.3}{mm}$\footnote{In Sec.~\ref{sec:H_d} we vary the magnetic field and hence also the resonant wave number $k_q$. Accordingly the group velocity $v_{k_q}$ varies over time. Here, we take the minimal value of the group velocity $ v_{k_{q}} \approx \SI{0.3}{km /s }$ to estimate the coherence length $l_m$.}.

The magnetic moments of the spins interact with the magnetic field of the magnon modes. Under the rotating wave approximation, this interaction is given by the Hamiltonian~\cite{trifunovicLongDistanceEntanglementSpin2013,flebusEntanglingDistantSpin2019,bertelliMagneticResonanceImaging2020,neumanNanomagnonicCavitiesStrong2020,fukamiOpportunitiesLongRangeMagnonMediated2021,gonzalez-ballesteroQuantumInterfaceSpin2022,hetenyiLongdistanceCouplingSpin2022,karanikolasMagnonmediatedSpinEntanglement2022,fukamiMagnonmediatedQubitCoupling2024,bejaranoParametricMagnonTransduction2024,xueDirectionalEntanglementSpinorbit2025}
\begin{equation}
    \hat{H}_{Q,M} = \hbar\sum_{k,i}(g_{k,i}^{-} \hat{\sigma}^{-0}_{i} + g_{k,i}^{+}\hat{\sigma}^{+0}_{i})\hat{m}_{k} + \text{h.c},
\label{eq:H_Q_M}
\end{equation}
with coupling constant $g_{k,i}^{\pm}=|\gamma_{s}|\mu_{0}\delta\boldsymbol{H}_{k}(\boldsymbol{r}_{i})\cdot\boldsymbol{e}_{\mp}$.
This coupling falls exponentially in the wave number $g_{k,i}^{\pm}\propto e^{-|k|x_i}$~\cite{gonzalez-ballesteroQuantumInterfaceSpin2022,dolsSteadystateEntanglementSpin2026}.
Since the wave number $|k_+|$ which is resonant with the $\ket{0}\leftrightarrow\ket{+}$ transition of the NV center is larger than the wave number $|k_-|$ which is resonant with the $\ket{0}\leftrightarrow\ket{-}$ transition, one finds $g_{|k_{+}|,i}^{+} / g_{|k_{-}|,i}^{-} = 10^{-2}$.
Thus, we neglect the coupling of the magnons to the $\ket{+}$ state of the NV and take $\omega_{-}$ as our qubit frequency $\omega_q$ and consequently $k_-$ as our resonant wave number $k_q$.

By modulating the position $x_{i}(t)$ of qubit $i$ in time, the coupling $g_{k,i}^{-}(t)$ of the magnons to NV center $i$ is modulated accordingly, and hence the dissipative coupling $J_{i}(t)$.
Owing to the short wavelengths of magnons, typically in the micron regime, the variation of the NV-magnet distance $x_i(t)$ must also ensue on the scale of microns.
This implies that this protocol can be implemented without the need of bulky systems.
With $x_{i}(t) = d/2 + d_{0} + \delta x_{i}(t)$, where $d_{0}$ is a fixed distance between the magnetic surface and the NV center and where $\delta x_{i}(t)>0$ is the deviation from this distance, we rewrite the coupling such that $g_{k,i}^{-}(t)=\tilde{g}_{k,i}^{-}\exp[-|k|x_{i}(t)]$.
Then, the dissipative coupling can be written such that
\begin{equation}
    J_{i}(\delta x_{i} (t) )= J_{0}e^{-2|k_{q}|\delta x_{i}(t)},
\label{eq:J_x_t}
\end{equation}
where $J_{0}=|\tilde{g}_{i,k_{q}}^{-}|^{2}\exp[-|k_{q}|(d + 2d_{0})]L_y/v_{k_{q}}$.

\begin{table}[h]
\caption{Summary of the parameters of the NV centers~\cite{dohertyNitrogenvacancyColourCentre2013} and the YIG stripe~\cite{stancilSpinWavesTheory2009}.\label{tab:vals}}
\begin{ruledtabular}
\begin{tabular}{lc}
\textrm{Parameter}&\multicolumn{1}{c}{\textrm{Value}}\\
\colrule
NV zero-field splitting & $D_0 = 2\pi \times \SI{2.877}{GHz}$ \\
Gyromagnetic ratio & $\gamma_s = \SI{1.76e11}{T^{-1}s^{-1}}$ \\
YIG stripe thickness & $d=\SI{150}{nm}$ \\
YIG stripe length & $L_y = 100 d_0$ \\
YIG stripe width & $ L_z = 15 d_0$  \\
YIG exchange stiffness & $D_{\mathrm{ex}} = \SI{ 3.086e-16}{m^{2}}$ \\
YIG saturation magnetization & $M_s = \SI{1.39 e5}{A m^{-1}}$ \\
Minimal distance NV to YIG surface & $d_{\mathrm{min}} = \SI{50}{nm}$ \\
\end{tabular}
\end{ruledtabular}
\end{table}

In the next subsection~\ref{sec:d_1_2}, we modulate both distances $\delta x_{1,2} (t)$ such that the desired couplings $J_{1,2}(t)$ are obtained.
In practice, the modulation of the distances can be done mechanically using cantilevers~\cite{rondinNanoscaleMagneticField2012,maletinskyRobustScanningDiamond2012,sunMagneticDomainsDomain2021,hacheNanoscaleMappingMagnetic2025}, as depicted schematically in Fig.~\eqref{fig:setup}.

\subsection{Varying the distances $\delta x_1$ and $\delta x_2$}\label{sec:d_1_2}

\begin{figure}[ht]
\includegraphics[width=\linewidth]{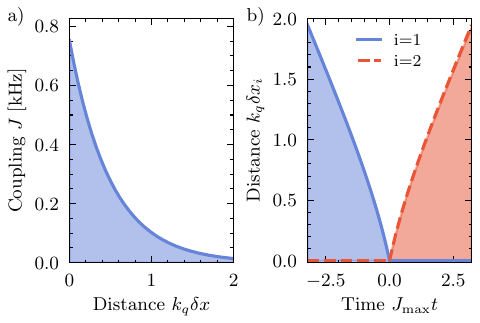}
\caption{\label{fig:d_J} a) Coupling strength $J$ as a function of the distance $\delta x$ normalized by the resonant wave number $k_q$. b) Distance $\delta x_{i}$ of qubit $i$ normalized by the resonant wave number $k_q$ as a function of the time $t$ normalized by the coupling $J_{\mathrm{max}}$.}
\end{figure}

Since the coupling strength $J_0$ controls the protocol time, we position of the qubits to be as close to the magnetic surface as possible in order to obtain the largest coupling.
Thus, we take $d_0 = d_\mathrm{min}$, where $d_\mathrm{min}$ is the minimal allowed distance between the NV center and the magnet, giving rise to the maximal coupling $ J_\mathrm{max} = J_0 (d_0 = d_{\mathrm{min}})$.
We show the dependence of the coupling given in Eq.~(\ref{eq:J_x_t}) on the variation of the distance $\delta x$ in Fig.~\ref{fig:d_J}a), using the parameters defined in Tab.~\ref{tab:vals}.

The required $\delta x_{1,2} (t)$ to obtain the effective couplings $J_i(t)$ can be obtained by setting the desired coupling of Eq.~(\ref{eq:J_t}) equal to the coupling $J_{i}(\delta x_{i} (t) )$ of Eq.~(\ref{eq:J_x_t}),
\begin{equation}
   \delta x_{1}(t) =\begin{cases}
\frac{1}{2|k_{q}|}\ln\left(2e^{- J_{\mathrm{max}}t}-1\right), & t<0\\
0, & t\geq0
\end{cases}
\label{eq:delta_x_1}
\end{equation}
and
\begin{equation}
\delta x_{2}(t) = 
\begin{cases}
0, & t<0\\
\frac{1}{2|k_{q}|}\ln\left(2e^{ J_{\mathrm{max}}t}-1\right), & t\geq0.
\end{cases}
\label{eq:delta_x_2}
\end{equation}
The time dependence of $\delta x_{1,2} (t)$ is shown in Fig.~\ref{fig:d_J}b).

In the field of magnetometry, experiments have shown that the minimal NV-probe distance can be tens of nanometers~\cite{rondinNanoscaleMagneticField2012,maletinskyRobustScanningDiamond2012,sunMagneticDomainsDomain2021,hacheNanoscaleMappingMagnetic2025} and hence we take $d_{\mathrm{min}}=\SI{50}{nm}$.
Using the parameters in Table~\ref{tab:vals}, we find $J_{\mathrm{max}} = \SI{0.8}{kHz}$.
To obey the phase matching required for the protocol such that the local phase of the input state is preserved, the NV centers should be positioned at a distance $y_{2,1} =  \pi/k_q\times (1+2n) = \SI{0.4}{\micro m} \times (1+ 2n)$, with integer $n$.
We note that by choosing a different phase $k_qy_{2,1}$, a different local phase of the final state $\alpha \ket{0}_{2} - \beta e^{-i k_q y_{2,1}} \ket{1}_2$ is achieved (see Eq.~(\ref{eq:c_f}) in Sec.~\ref{sec:prot}).
Changing this phase can be done by putting the qubits at a different distance $y_{2,1}$ or by tuning the dispersion of the magnons through the magnetic field such that the resonant wave vector $k_q$ is different.
We note that $\tau_g = \mathcal{O}(1/J_{\mathrm{max}})$ and hence $\tau_g \gg \tau_m$ holds.
The qubit's positions are varied over about $\SI{250}{nm}$ on a timescale of $\SI{8}{ms}$, see Fig.~\ref{fig:d_J}b).
The amplitude of the external field amounts $\mu_0 H_0 = \SI{8}{mT}$.

Considering that the dephasing time $T_\phi$ of NV centers is typically much faster than the lifetime $T_1$~\cite{rondinMagnetometryNitrogenvacancyDefects2014}, we assume that the lifetime is of negligible influence on the protocol with respect to dephasing.
Then, the benchmark of Fig.~\ref{fig:protocol}c) gives that a dephasing time of $T_\phi = 20/J$ is required to achieve a fidelity $F>0.95$. Using the coupling $J_{\mathrm{max}} = \SI{0.8}{kHz}$ found above, this corresponds to $T_\phi = \SI{25}{ms}$.

\subsection{Varying the magnetic field $H_0$ and the distance $\delta x_2$}\label{sec:H_d}

An experimental challenge of the realization of the setup in Sec.~\ref{sec:d_1_2} is to engineer two positioning devices, e.g. cantilevers, with NV centers within a distance in the micron regime.
To circumvent the necessity of having two moving NV centers, we propose to move only one qubit, while exploiting another tuning parameter: the external magnetic field $H_0$.
In this way, our scheme can be implemented on existing scanning NV magnetometer setups.

\begin{figure}[ht]
\includegraphics[width=\linewidth]{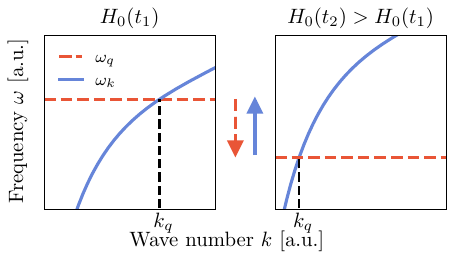}
\caption{\label{fig:dispersion_H_t} The qubit frequency $\omega_q$ and the magnon dispersion $\omega_k$ can be tuned through the magnetic field $H_0(t)$, such that the resonant wave number $k_q$ can be varied by tuning the magnetic field.
The arrows indicate the direction in which the frequencies change when increasing $H_0(t)$.}
\end{figure}

\begin{figure}[t]
\includegraphics[width=\linewidth]{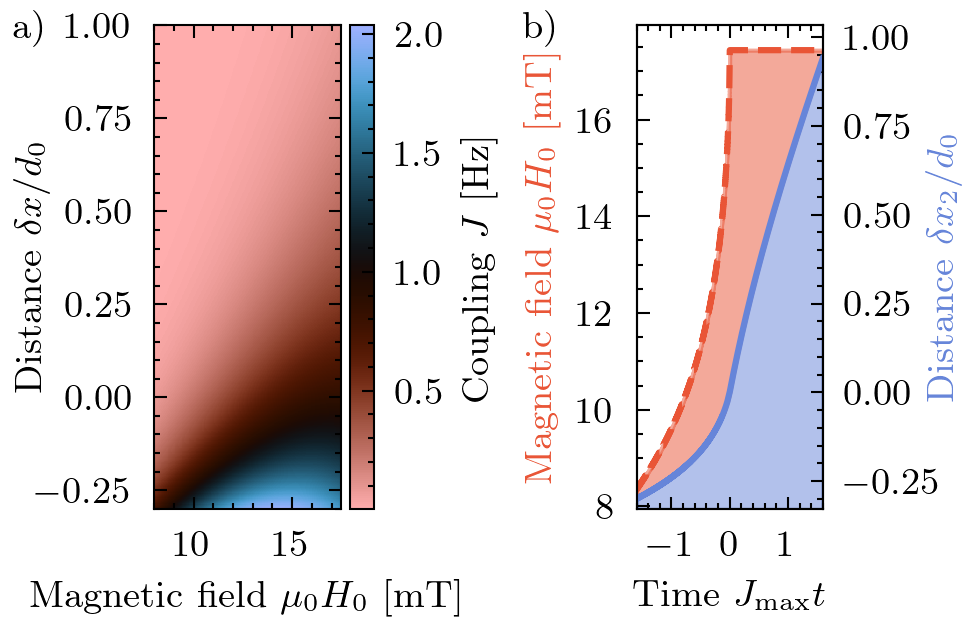}
\caption{\label{fig:H_d_J} a) Coupling strength $J$ as a function of the magnetic field $\mu_0 H_0$ and the distance $\delta x$ normalized by $d_0$. b) The magnetic field $\mu_0 H_0$ in blue and distance $\delta x_2$ normalized by $d_0$ as a function of the time $t$ normalized by the coupling $J_{\mathrm{max}}$.}
\end{figure}

The external magnetic field influences both the magnon dispersion as well as the NV frequency.
By changing the amplitude of the magnetic field, the resonant wave number $k_q(t)$ can be varied (see Fig.~\ref{fig:dispersion_H_t} and App.~\ref{app:EOM} for the derivation) and hence the dissipative coupling $J$.
We show the dependence of the dissipative coupling on the variation of the magnetic field and distance in Fig.~\ref{fig:H_d_J}a).
We take the first qubit as stationary $\delta x_1 (t) = 0$, while the position of the second qubit is varied in time.
Thus, one finds $J_1(t)=J_{0}(t)$ according to Eq.~(\ref{eq:J_x_t}).
The only tuning parameter for the coupling $J_1(t)$ is the magnetic field, see Fig.~\ref{fig:H_d_J}a).
Thus, one needs to find the magnetic field such that the desired coupling $J_1(t)$ of Eq.~(\ref{eq:J_t}) is implemented.
One finds the magnetic field modulation as a function of time as presented in Fig.~\ref{fig:H_d_J}b).
In turn, the distance $\delta x_{2}(t)$ is varied such that also the coupling $J_2(t)$ is realized.
With Eq.~(\ref{eq:J_x_t}) one finds $J_{2}(t)=J_{1}(t)e^{-2|k_{q}(t)|\delta x_{2}(t)}$.
Using Eq.~(\ref{eq:J_1_2}), this relation can be inverted to obtain $\delta x_2(t)$ as a function of $k_q(t)$ and the function $f(t)$ (see Sec.~\ref{sec:prot}), giving rise to
\begin{equation}
    \delta x_{2}(t)=-\frac{1}{2|k_{q}(t)|}\ln\left(\frac{1-f(t)}{f(t)}\right).
\end{equation}
The dependence of $\delta x_{2}(t)$ on the resonant wave number $k_{q}(t)$ implies that the second qubit's distance has to be varied at length scales of the order of magnitude of the magnet thickness $d$.
Therefore, the distance $d_{0}$ has to be sufficiently larger than $d_\mathrm{min}$, such that the distance between the NV and the magnetic surface is not lower than minimal distance $d_{\mathrm{min}}$ allowed between the NV and the magnet, i.e. $\mathrm{min}_{t}[\delta x_{2}(t)]+d_{0}\geq d_{\mathrm{min}}$.
As a consequence of this larger distance $d_0$, the NV-magnet coupling is lower than in the previous section where we had $d_0=d_\mathrm{min}$, implying a higher protocol time
$t_{p}$ than in the previous section.

The desired couplings $J_{1(2)}(t)$ start (stop) at a value close to zero and stop (start) at its maximum value, see Fig.~\ref{fig:protocol}a).
Thus, the ratio of maximum and minimum of the achievable coupling $J_{\mathrm{max}}/J_{\mathrm{min}}$ has to be high enough such that the minimal coupling $J_{\mathrm{min}}$ is sufficiently low with respect to the maximal coupling $J_{\mathrm{max}}$ in order to implement the protocol.
This ratio increases the higher one takes the distance $d_{0}$.
However, a higher $d_{0}$ comes with a lower maximal coupling $J_{\mathrm{max}}$.
In order to keep a reasonable coupling $J_{\mathrm{max}}$, which controls the protocol time $t_{p}$ while achieving a sufficient ratio $J_{\mathrm{max}}/J_{\mathrm{min}}$, we take $d_{0}=\SI{0.5}{\micro m}$ and $J_{\mathrm{max}}t_{f}=1.6$
(corresponding to $F_{T}=0.95$ as per Eq.~(\ref{eq:F_t_f})), giving $J_{\mathrm{max}}=\SI{1}{Hz}$. This corresponds to a minimal required NV dephasing time of $T_\phi = \SI{20}{s}$ to achieve a fidelity $F>0.87$. The magnetic field $H_0(t)$ and distance $\delta x_2 (t)$ are shown in Fig.~\ref{fig:H_d_J}b). 

\begin{figure}[h]
\includegraphics[width=\linewidth]{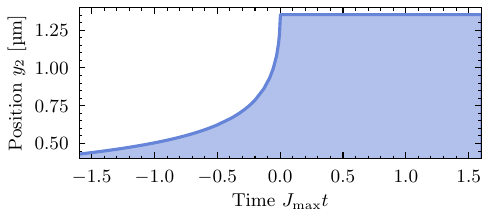}
\caption{\label{fig:Jt_y2} Position $y_2$ as a function of time $t$ normalized by the coupling $J_{\mathrm{max}}$.}
\end{figure}

Since the phase condition $k_{q}(t)y_{2,1}=\pi+2\pi n$ is required (see Sec.~\ref{sec:prot}) and we vary the resonant wave vector $k_{q}(t),$ the distance $y_{2,1}$ requires modulation as well.
We keep the position of the first qubit fixed at $y_{1}=0$.
In order to keep $y_2(t)$ continuous, the second qubit's position requires to be varied such that $y_{2}(t)=n\pi / k_q(t)$.
The position $y_2(t)$ for $n=1$ is shown in Fig.~\ref{fig:Jt_y2}.

The time modulation of the magnetic field is on a scale $\tau_{H_0} = \mathcal{O}(1/J_{\mathrm{max}})$, see Fig.~\ref{fig:H_d_J}b).
Since the variation of the qubit and magnon frequency are also on the time scale $\tau_{\omega_q},\tau_{\omega_k}\sim\tau_{H_0}$, and the coupling varies on a scale $\tau_g = \mathcal{O} (1/J_{\mathrm{max}})$, one finds $\tau_{\omega_q},\tau_{\omega_k} \gg \tau_m$.
The position of the second qubit is varied over about $\SI{150}{nm}$ and the magnetic field over $\SI{9}{mT} / \mu_0$ on a timescale of $\SI{3.2}{s}$, see Fig.~\ref{fig:H_d_J}b).

By changing the frequency of the NV centers by e.g. applying a local magnetic field, their interaction can be tuned between off-resonance and resonance.
Before the start of the protocol, the NV centers are off-resonant such that they do not couple through the bath.
At $t=t_i$ the local fields are turned off to start their interaction. By detuning the qubits again at $t=t_f$, the coupling is off-resonant and hence finalizes the protocol.

\section{Conclusions}\label{sec:con}

We proposed a protocol for the deterministic transfer of an arbitrary quantum state between distant spin qubits mediated by chiral magnons.
The coupling between the spin qubits and the chiral magnon bath is assumed to be tunable, and we showed that, by integrating out the magnons within the Born-Markov approximation, a time-dependent unidirectional qubit-qubit interaction is obtained.
By tuning the qubit-qubit coupling such that i) the two-qubit state remains a dark state and ii) the excitation of the first qubit is transferred to the second while maintaining the local phase, a state-transfer protocol without dissipation into the bath is realized.
We benchmarked this protocol with respect to the intrinsic lifetime and dephasing time of the qubits and with respect to the temperature required.

We numerically investigated the feasibility of the protocol on a hybrid system where two NV centers are coupled to the nonreciprocal and chiral surface modes of a magnetic YIG stripe. 
We proposed two implementations for tuning the couplings and therefore realize the protocol.
In the first implementation, the external bias magnetic field remains constant and the distances of the NV centers to the magnet are varied.
We showed that in order to achieve a state-transfer fidelity $\gtrsim 0.95$, a dephasing time $T_{\phi} \gtrsim \SI{25}{ms}$ is required.
While such dephasing times are challenging to achieve, dynamical decoupling techniques~\cite{violaDynamicalDecouplingOpen1999,caiRobustDynamicalDecoupling2012,frydrychSelectiveDynamicalDecoupling2014,arrazolaEngineeringProtectedCavityQED2025a} have been used to successfully improve $T_{\phi}$ in NV-based systems~\cite{bar-gillSolidstateElectronicSpin2013,abobeihOnesecondCoherenceSingle2018,yamamotoTenSecondElectronSpinCoherence2026}. 
We note that unidirectional coupling schemes have shown compatibility with a continuous drive applied to the qubits~\cite{stannigelDrivendissipativePreparationEntangled2012,pichlerQuantumOpticsChiral2015,dolsSteadystateEntanglementSpin2026}, i.e. the central ingredient in continuous dynamical decoupling~\cite{caiRobustDynamicalDecoupling2012}.
In the second implementation, the magnetic field and the position of the second qubit are varied, while the position of the first qubit is kept constant.
While such an implementation might have experimental advantages, as only one NV needs to be moved, it yields a longer protocol time as consequence of the larger distance between the first qubit and the magnet.
Consequently, longer dephasing times are required. For instance, to achieve a fidelity $ \gtrsim 0.87$, the required dephasing time is $T_\phi \gtrsim \SI{20}{s}$, significantly larger than the first scheme.
We benchmarked a temperature $\mathcal{T} \lesssim \SI{25}{mK}$ to work within the zero-temperature model.

The two central ingredients of the protocol in this work are the dark-state condition, which prevents loss into the magnon bath, and the time modulation of the coupling of the spin qubits to the magnon bath.
The tuning knobs used to modulate the coupling are the distance of the NV center to the magnetic strip and the external magnetic field, enabling coupling engineering and dispersion engineering, respectively.
Natural extensions could involve considering different magnet geometries~\cite{otaloraCurvatureInducedAsymmetricSpinWave2016,gallardoReconfigurableSpinWaveNonreciprocity2019} for improving scalability, or including topological magnons as mediators~\cite{hetenyiLongdistanceCouplingSpin2022}.
Our results therefore provide a general framework for
using chiral and nonreciprocal magnons as tunable quantum channels in
spin-based quantum technologies.

\begin{acknowledgments}

M.D., M.C., and S.V.K. acknowledge financial support by the Federal Ministry of Research, Technology and Space (BMFTR) project QECHQS (Grant No. 16KIS1590K).

\end{acknowledgments}

\section*{Author declarations}

\subsection*{Conflict of Interest}
The authors have no conflicts to disclose.

\subsection*{Author Contributions}

\textbf{M. Dols}: Conceptualization (equal); Formal analysis; Investigation; Methodology (lead); Visualization; Writing – original draft (lead).
\textbf{C. Gonzalez-Ballestero}: Conceptualization (equal); Methodology (supporting); Writing – original draft (supporting).
\textbf{M. Cherkasskii}: Conceptualization (equal); Methodology (supporting); Writing – original draft (supporting).
\textbf{C.L. Degen}: Conceptualization (equal);  Writing – original draft (supporting).
\textbf{V.A.S.V. Bittencourt}: Conceptualization (equal); Methodology (supporting); Funding acquisition (supporting); Supervision (equal);  Writing – original draft (supporting).
\textbf{S. Viola Kusminskiy}: Conceptualization (equal); Methodology (supporting); Funding acquisition (lead); Project administration; Supervision (equal);  Writing – original draft (supporting).

\section*{Data Availability Statement}
The data that support the findings of this study are available
from the corresponding author upon reasonable request.

\appendix

\section{Effective qubit-qubit equation of motion}\label{app:EOM}

Here, we derive the effective qubit-qubit dynamics for finite temperatures.
Besides the time dependency of the magnon-qubit coupling $g_{k,i}(t)$ as introduced in Sec.~\ref{sec:model}, we allow for a time dependency in the qubit frequency $\omega_q(t)$, which is assumed to be identical for both qubits, and in the magnon dispersion $\omega_{k}(t)$.
The results of this appendix can hence be applied to Sec.~\ref{sec:model} and Sec.~\ref{sec:H_d} in the respective regime of time-dependent parameters.
For treating time-dependent systems coupled to a bath we refer to Refs.~\onlinecite{rousochatzakisMasterEquationsPulsed2005,breuerTheoryOpenQuantum2007,caiDynamicEntanglementOscillating2010,stannigelOptomechanicalTransducersQuantuminformation2011a}.
To obtain the equation of motion describing unidirectional qubit-qubit coupling we follow closely the derivation presented in Ref.~\onlinecite{dolsSteadystateEntanglementSpin2026}.
We show that for low temperatures $k_B \mathcal{T} \ll \hbar\omega_q $ and time-independent magnon dispersion and qubit frequency, we retrieve the findings presented in Sec.~\ref{sec:model}.
We use the time dependence of the magnon dispersion and qubit frequency in Sec.~\ref{sec:H_d}.

The von Neumann equation dictates the time propagation of the density operator $\hat{\rho}_{\mathrm{tot}}^{I}$ describing the total system, which we take in the interaction picture.
This equation can be integrated formally, giving rise to~\cite{breuerTheoryOpenQuantum2007}
\begin{equation}
\begin{aligned}
\frac{\mathrm{d}}{\mathrm{d}t}\hat{\rho}_{\mathrm{tot}}^{I}(t) & =-\frac{i}{\hbar}\left[\hat{H}_{\mathrm{int}}^{I}(t),\hat{\rho}_{\mathrm{tot}}^{I}(0)\right] \\ & \quad-\frac{1}{\hbar^{2}}\int_{0}^{t}\mathrm{d}\tilde{t}\left[\hat{H}_{\mathrm{int}}^{I}(t),\left[\hat{H}_{\mathrm{int}}^{I}(\tilde{t}),\hat{\rho}_{\mathrm{tot}}^{I}(\tilde{t})\right]\right]\,.
\end{aligned}
\label{eq:EOM_von_N}
\end{equation}
Here, $\hat{H}_{\mathrm{int}}^{I}(t)=\hat{U}_{I}(t) \hat{H}_{\mathrm{int}} \hat{U}_{I}^{\dagger}(t)$ is the interaction Hamiltonian in the interaction picture with $\hat{H}_{\mathrm{int}}$ given in Eq.~(\ref{eq:H_int}).
$\hat{U}_{I}(t) =\exp\left[i\int_{0}^{t}\left(\hat{H}_{Q}(\tilde{t}) + \hat{H}_{M}(\tilde{t})\right)\mathrm{d}\tilde{t} / \hbar\right]$ is a unitary transformation with the Hamiltonians $\hat{H}_{Q}(t)$ and $\hat{H}_{M}(t)$ given in Eqs.~(\ref{eq:H_Q}) and~(\ref{eq:H_M}), respectively.
We note that no time-ordering operator is required because the free Hamiltonian $\hat{H}_Q(t)+\hat{H}_M(t)$ commutes with itself at different times.
This gives rise to
\begin{equation}
    \hat{H}_{\mathrm{int}}^{I}(t)=\sum_{i,k}\hbar\left(g_{k,i}(t)\hat{\sigma}_{i}^{+}(t)\hat{m}_{k}(t)+g_{k,i}^{*}(t)\hat{m}_{k}^{\dagger}(t)\hat{\sigma}_{i}^{-}(t)\right),
\end{equation}
where $\hat{m}_{k}(t)=\exp\left[-i\varphi_{k}(t)\right]\hat{m}_{k}(0)$ and $\hat{\sigma}_{i}^{-}(t)=\exp\left[-i\varphi_{q}(t)\right]\hat{\sigma}_{i}^{-}(0)$, with $\varphi_{k}(t)=\int_{0}^{t}\omega_{k}(\tilde{t})\mathrm{d}\tilde{t}$ and $\varphi_{q}(t)=\int_{0}^{t}\omega_{q}(\tilde{t})\mathrm{d}\tilde{t}$.

We employ the Born Approximation, which states that the density matrix
$\hat{\rho}_{\mathrm{tot}}^{I}$ can be separated such that $\hat{\rho}_{\mathrm{tot}}^{I}(t)=\hat{\rho}^{I}(t)\otimes\hat{\rho}_{\mathrm{th}}$,
where $\hat{\rho}^{I}(t)$ describes the spin qubits and $\hat{\rho}_{\mathrm{th}}$
is the thermal state describing the magnon bath.
This approximation is valid provided that the interaction between the spin qubits and the magnons is weak.
Tracing over the magnonic degrees of freedom in Eq.~(\ref{eq:EOM_von_N}),
the first term gives $\mathrm{Tr}_{M}\left\{ \left[\hat{H}_{\mathrm{int}}^{I}(t),\hat{\rho}^{I}(0)\otimes\hat{\rho}_{\mathrm{th}}\right]\right\} =0$, because the interaction $\hat{H}_{\mathrm{int}}$ is linear in the magnon operator.
We define the time difference $\tau=t-\tilde{t}$, substitute this in Eq.~(\ref{eq:EOM_von_N}) and rewrite the second term such that
\begin{widetext}
\begin{equation}
\begin{aligned}
    \frac{\mathrm{d}}{\mathrm{d}t}\hat{\rho}^{I}(t) = -\sum_{i,j}\int_{0}^{t}\mathrm{d}\tau  & \Big(\mathcal{K}_{i,j}^{\downarrow}(t,t-\tau)\left[\hat{\sigma}_{i}^{+}(t),\hat{\sigma}_{j}^{-}(t-\tau)\hat{\rho}^{I}(t-\tau)\right]+\mathcal{K}_{j,i}^{\downarrow}(t-\tau,t)\left[\hat{\rho}^{I}(t-\tau)\hat{\sigma}_{j}^{+}(t-\tau),\hat{\sigma}_{i}^{-}(t)\right]\\ &+\mathcal{K}_{i,j}^{\uparrow}(t,t-\tau)\left[\hat{\rho}^{I}(t-\tau)\hat{\sigma}_{j}^{-}(t-\tau),\hat{\sigma}_{i}^{+}(t)\right]+ \mathcal{K}_{j,i}^{\uparrow}(t-\tau,t)\left[\hat{\sigma}_{i}^{-}(t),\hat{\sigma}_{j}^{+}(t-\tau)\hat{\rho}^{I}(t-\tau)\right] ] \Big),
\end{aligned}
\label{eq:EOM_kernel}
\end{equation}
where we defined the kernels
\begin{equation}
    \mathcal{K}_{i,j}^{\downarrow}(t,t-\tau)=\sum_{k,k'}g_{k,i}(t)g_{k',j}^{*}(t-\tau)\langle\hat{m}_{k}(t)\hat{m}_{k'}^{\dagger}(t-\tau)\rangle, \quad \mathcal{K}_{i,j}^{\uparrow}(t,t-\tau)=\sum_{k,k'}g_{k,i}(t)g_{k',j}^{*}(t-\tau)\langle\hat{m}_{k'}^{\dagger}(t-\tau)\hat{m}_{k}(t)\rangle
\end{equation}
\end{widetext}
and used $\langle\hat{m}_{k}(t)\hat{m}_{k'}(\tilde{t})\rangle=\langle\hat{m}_{k}^{\dagger}(t)\hat{m}_{k'}^{\dagger}(\tilde{t})\rangle=0$, where the expectation value of the magnon operators is taken with respect to the thermal state $\hat{\rho}_{\mathrm{th}}$.
As shown below, the kernels $\mathcal{K}_{i,j}^{\downarrow}$ and $\mathcal{K}_{i,j}^{\uparrow}$ are associated with qubit emission into and absorption from the magnon bath, respectively.
We apply the Markov approximation, which assumes that the correlation time of the magnon bath $\tau_{m}$ is much shorter than of the qubit $\tau_{q}$.
Due to the weak-coupling assumption and $\tau_m\ll\tau_q$, the state of the qubits does not vary on the scale of the bath-correlation time $\tau_{m}$, one can approximate $\hat{\rho}^{I}(t-\tau)\approx\hat{\rho}^{I}(t)$.
In the kernels there are correlation functions of the bath, e.g., $\langle\hat{m}_{k}^{\dagger}(t)\hat{m}_{k'}(t-\tau)\rangle\propto \exp(-\kappa_m \tau)$, where $\kappa_m = 1/\tau_m$.
These correlations vanish for times $\tau\gg\tau_{m}$.
Expanding the coupling $g_{k,i}(t-\tau)$ in $\tau$ around the slow time $t$ gives
\begin{equation}
    g_{k,i}(t-\tau) = g_{k,i}(t) - \dot{g}_{k,i}(t) \tau + \mathcal{O}(\tau^2).
\end{equation}
Thus, if $|\dot{g}_{k,i}(t) \tau_m  / g_{k,i}(t)| \ll 1$, one can approximate $g_{k,i}(t-\tau) = g_{k,i}(t)$, since $\tau$ is on the time scale given by $\tau_m$.
This leads to the condition $\tau_m \ll \tau_g$, where $\tau_g \sim  |g_{k,i} / \dot{g}_{k,i}|$.
In Sec.~\ref{sec:H_d}, the dispersion of the magnon and the qubit frequency are varied in time.
Using the unitary $\hat{U}_{I}(t)$ one finds
\begin{equation}
\begin{aligned}
    \langle\hat{m}_{k}(t)\hat{m}_{k'}^{\dagger}(t-\tau)\rangle = & \exp\{i[\varphi_{k}(t-\tau)-\varphi_{k}(t)] - \kappa_m\tau\} \\ & \times \left(\bar{n}_{k}(t)+1\right)\delta_{k,k'}, 
\end{aligned}
\end{equation}
where $\bar{n}_{k}(t)=1/(\exp(\hbar\omega_{k}(t)/k_{B}\mathcal{T})-1)$ is the thermal occupation number.
With
\begin{equation}
    \varphi_{k}(t-\tau)-\varphi_{k}(t) = -\int_{0}^{\tau}\omega_{k}(t-\tilde{t})\mathrm{d}\tilde{t},
\end{equation}
one can expand the magnon dispersion $\omega_{k}(t-\tilde{t}) = \omega_{k}(t)-\dot{\omega}_{k}(t)\tilde{t}+\mathcal{O}\left(\tilde{t}^{2}\right)$, 
giving
\begin{equation}
    \varphi_{k}(t-\tau)-\varphi_{k}(t) = -\omega_{k}(t)\tau+\frac{1}{2}\dot{\omega}_{k}(t)\tau^{2}+\mathcal{O}(\tau^{3}).
\label{eq:phi_k}
\end{equation}
Thus, for $\tau_m \ll \tau_{\omega_{k}}$, where $\tau_{\omega_{k}} \sim | \omega_{k}/ \dot{\omega}_{k} |$ we can take 
$\varphi_{k}(t-\tau)-\varphi_{k}(t) = -\omega_{k}(t)\tau$.
The magnon bath thermalizes at a time scale given by $\tau_m$.
Since this time scale is much shorter than the time scale $\tau_{\omega_{k}}$ at which we vary the frequency of the magnon, we can use the thermal occupation number $\bar{n}_{k}(t)$ evaluated at time $t$ as given above.
This condition has a different physical meaning from the Markov approximation, although both are controlled by the same time $\tau_m$ in the present model.
Accordingly, we can rewrite the first kernel to
\begin{equation}
    \mathcal{K}_{i,j}^{\downarrow}(t,t-\tau)=\sum_{k}g_{k,i}(t)g_{k,j}^{*}(t)\left(\bar{n}_{k}(t)+1\right)e^{-[i\omega_{k}(t)+\kappa_m]\tau}.
\end{equation}
In analogy, the second kernel can be rewritten to
\begin{equation}
    \mathcal{K}_{i,j}^{\uparrow}(t,t-\tau)=\sum_{k}g_{k,i}(t)g_{k,j}^{*}(t)\bar{n}_{k}(t)e^{-[i\omega_{k}(t)+\kappa_m]\tau},
\end{equation}
where we used $\langle\hat{m}_{k'}^{\dagger}(t)\hat{m}_{k}(t)\rangle=\bar{n}_{k}(t)\delta_{k,k'}$.
Extending the upper limit of the integral in Eq.~(\ref{eq:EOM_kernel}) to $\infty$~\cite{breuerTheoryOpenQuantum2007} gives rise to
\begin{widetext}
\begin{equation}
\frac{\mathrm{d}}{\mathrm{d}t}\hat{\rho}^{I}(t)=-\sum_{i,j}\left(J_{i,j}^{\downarrow}(t)\left[\hat{\sigma}_{i}^{+},\hat{\sigma}_{j}^{-}\hat{\rho}^{I}(t)\right]+\left(J_{i,j}^{\downarrow}(t)\right)^{*}\left[\hat{\rho}^{I}(t)\hat{\sigma}_{j}^{+},\hat{\sigma}_{i}^{-}\right]+J_{i,j}^{\uparrow}(t)\left[\hat{\rho}^{I}(t)\hat{\sigma}_{j}^{-},\hat{\sigma}_{i}^{+}\right]+\left(J_{i,j}^{\uparrow}(t)\right)^{*}\left[\hat{\sigma}_{i}^{-},\hat{\sigma}_{j}^{+}\hat{\rho}^{I}(t)\right]\right),
\label{eq:EOM_BM}
\end{equation}
where we defined the coupling strengths
\begin{equation}
    J_{i,j}^{\downarrow}(t) = \sum_{k}\int_{0}^{\infty}\mathrm{d}\tau g_{k,i}(t)g_{k,j}^{*}(t)\left(\bar{n}_{k}(t)+1\right)e^{-[i\delta_{k}(t)+\kappa_m]\tau}, \quad J_{i,j}^{\uparrow}(t)= \sum_{k}\int_{0}^{\infty}\mathrm{d}\tau g_{k,i}(t)g_{k,j}^{*}(t)\bar{n}_{k}(t)e^{-[i\delta_{k}(t) + \kappa_m]\tau}.
\end{equation}
with the detuning $\delta_{k}(t) = \omega_{k}(t)-\omega_{q}(t)$.
Here, we approximated $\varphi_{q}(t-\tau)-\varphi_{q}(t) = -\omega_{q}(t)\tau$, in analogy with Eq.~(\ref{eq:phi_k}), valid for $\tau_m \ll \tau_{\omega_q}$, where $\tau_{\omega_q} \sim | \omega_{q}(t) / \dot{\omega}_{q}(t) |$.

Eq.~(\ref{eq:EOM_BM}) is the time-dependent equivalent of Eq.~(B3) in Ref.~\onlinecite{dolsSteadystateEntanglementSpin2026}.
We follow the steps provided in that work to obtain the effective coupling.
Performing the integral over $\tau$ gives rise to
\begin{equation}
    J_{i,j}^{\downarrow}(t)=\frac{L}{2\pi}\int\mathrm{d}kg_{k,i}(t)g_{k,j}^{*}(t)\left(\bar{n}_{k}(t)+1\right)\frac{1}{i\delta_{k}(t)+\kappa_{m}}.
\end{equation}
Here, $L$ is the length of the magnetic stripe along the magnon-propagation direction, which is the direction along which the two NV centers are separated.
We linearize $\delta_{k}(t)=v_{k_{q}}(t)[k- k_{q}(t)]$, where we only consider one sign of the resonant wave number (so $k_q(t)>0$), since the coupling is limited to positive wave numbers due to $g_{k<0,i}(t) = 0$.
Using $g_{k,i}(t)=|g_{k,i}(t)|e^{ikr_{i}}$ gives rise to
\begin{equation}
    J_{i,j}^{\downarrow}(t)=-i\frac{L}{2\pi}\int\mathrm{d}k \frac{|g_{k,i}(t)||g_{k,j}(t)|\left(\bar{n}_{k}(t)+1\right)}{v_{k_{q}}(t)}\frac{e^{ikr_{i,j}}}{k- [k_{q}(t)+i\kappa_{m}/v_{k_{q}}(t)]}
\end{equation}
Using the residue theorem and the wide-band approximation, valid for $\kappa_{m}/v_{k_{q}}(t) \ll k_q(t)$\footnote{With the values presented in Sec.~\ref{sec:imp}, one finds $\kappa_{m}/[v_{k_{q}}(t)k_q(t)] = \mathcal{O}(10^{-4})$.}, one finds $J_{i,j}^{\downarrow}(t) = \left(\bar{n}_{k_{q}}(t)+1\right)J_{i,j}(t)$, where
\begin{equation}
    J_{i,j}(t)=\frac{1}{2}\sqrt{J_{i}(t)J_{j}(t)}e^{[ik_{q}(t)-\alpha(t)]r_{i,j}}\left(1+\mathrm{sgn}\left(r_{i,j}\right)\right),
\end{equation}
with effective qubit-bath coupling $J_{i}(t)=|g_{i,k_{q}}(t)|^{2}L/v_{k_q}(t)$ and $\alpha(t)=1/l_m(t)$, where $l_m(t) = v_{k_{q}}(t) \tau_m$ is the correlation length.
Proceeding analogously to obtain the coupling $J_{i,j}^{\uparrow}(t)$, one finds $J_{i,j}^{\uparrow}(t) =\bar{n}_{k_{q}}(t)J_{i,j}(t)$.
For the state-transfer protocol considered in the main text, we assume that the qubit separation is much smaller than the magnon coherence length $ r_{i,j}\ll l_m(t)$ for all times during the protocol.
We therefore set $\exp(-\alpha(t)r_{i,j})=1$ in the following and omit propagation attenuation.
By plugging the results for $J_{i,j}^{\downarrow}(t)$ and $J_{i,j}^{\uparrow}(t)$ into Eq.~(\ref{eq:EOM_BM}), one obtains the master equation for the two-qubit operator $\hat{\rho}$ in the laboratory frame~\cite{dolsSteadystateEntanglementSpin2026}
\begin{equation}
    \frac{\mathrm{d}}{\mathrm{d}t}\hat{\rho}(t) = -\frac{i}{\hbar}\left([\hat{H}_{Q},\hat{\rho}] + \hat{H}_{\mathrm{eff},\mathcal{T}}(t)\hat{\rho}-\hat{\rho}\hat{H}_{\mathrm{eff},\mathcal{T}}^{\dagger}(t)\right)+\left(\bar{n}_{k_{q}}(t)+1\right)\hat{L}(t)\hat{\rho}\hat{L}^{\dagger}(t)+\bar{n}_{k_{q}}(t)\hat{L}^{\dagger}(t)\hat{\rho}\hat{L}(t),
\label{eq:Liou_T}
\end{equation}
\end{widetext}
where $\hat{H}_{\mathrm{eff},\mathcal{T}}(t)$ is the effective Hamiltonian for finite temperatures
\begin{equation}
    \hat{H}_{\mathrm{eff},\mathcal{T}}(t)=(1+\bar{n}_{k_{q}}(t))\hat{H}_{\mathrm{eff}}(t)+\bar{n}_{k_{q}}(t)\hat{H}_{\mathrm{abs}}(t).
\end{equation}
The first Hamiltonian reads
\begin{equation}
    \hat{H}_{\mathrm{eff}}=-i\hbar\sum_{i,j=1,2}J_{i,j}(t)\hat{\sigma}_{i}^{+}\hat{\sigma}_{j}^{-},
\label{eq:H_eff}
\end{equation}
Plugging in the definition of $J_{i,j}(t)$ into the Hamiltonian $\hat{H}_{\mathrm{eff}}$ for $i=j$ and $i\neq j$ gives the Hamiltonian of Eq.~(\ref{eq:H_loc}) and Eq.~(\ref{eq:H_uni}), respectively. 
The second Hamiltonian describes thermal absorption
\begin{equation}
    \hat{H}_{\mathrm{abs}}(t) = -i\hbar\sum_{i,j=1,2}J_{i,j}^{*}(t)\hat{\sigma}_{i}^{-}\hat{\sigma}_{j}^{+}.
\end{equation}
As one can see in Eq.~(\ref{eq:Liou_T}), by taking the low temperature limit $k_B \mathcal{T} \ll \hbar\omega_q $ and hence $\bar{n}_{k_q} \ll1$, the thermal effective Hamiltonian $\hat{H}_{\mathrm{eff},\mathcal{T}}(t)$ boils down to the Hamiltonian $\hat{H}_{\mathrm{eff}}(t)$ and the remaining jump term equals $\hat{L}(t)\hat{\rho}\hat{L}^{\dagger}(t)$. 
If one takes time-independent qubit and magnon frequencies, one retrieves the master equation as defined in Sec.~\ref{sec:model}.

\section{Derivation of time dependence of the effective couplings}\label{app:pulses}

In this Appendix, we derive the effective couplings $J_{1,2}(t)$ as a function of $f(t)$ such that a dark state-transfer is implemented. Plugging Eq.~(\ref{eq:c_f}) into the dark-state condition~(\ref{eq:dark}) yields
\begin{equation}
    J_{1}(t) =J_{2}(t)\frac{f(t)}{1-f(t)}.
\label{eq:dark_J}
\end{equation}
Then, with the Schrödinger equation~(\ref{eq:c_10}) and with the ansatz in Eq.~(\ref{eq:c_f}) one obtains
\begin{equation}
    J_{1}(t)=\frac{\dot{f}(t)}{1-f(t)}
\end{equation}
and hence with the dark-state condition~(\ref{eq:dark_J})
\begin{equation}
    J_{2}(t)=\frac{\dot{f}(t)}{f(t)}.
\end{equation}

%
\section{Effective couplings dictated by $g(t)$}\label{app:pulse_g}

Here, we discuss the properties of the effective couplings yielded by $f(t) = 1 / (1+e^{-g(t)})$, see Eq.~(\ref{eq:J_1_2}), where $g(t)$ is given in Eq.~(\ref{eq:g_t}).
We choose the parameter $A$ such that at the final time $t_f$ the fidelity reaches the threshold $F_{T}$, corresponding to $F_{T}^{2}=f(t_{f})$ according to Eq.~(\ref{eq:F_t_f}) for $\alpha = 0$ and $\beta = 1$.
This leads to the relation $A=\ln\left(F_{T}^{2}/(1-F_{T}^{2})\right)$, which is given in the main text.
Now, we link $A$ and the time $t_{f}$ with the the maximum of the couplings $J$. It is sufficient to do this for the coupling $J_1(t)$, since the couplings $J_{1,2}(t)$ share their maximum value $J$. We define $x=\pi t/(2t_{f})$ such that the coupling can be rewritten to
\begin{equation}
    J_{1}(x) = \frac{2A}{t_{f}}j_{1}(x),
\label{eq:J_j_x}
\end{equation}
where $j_{1}(x)=\cos^{2}(x)/(1+e^{-g(x)})$. Requiring $j_1'(x) = 0$ gives rise to the relation
\begin{equation}
    \sin(2x) = \frac{4A}{\pi}\frac{\cos^{4}(x)}{1+\exp[{\frac{A}{\pi}\left(2x+\sin\left(2x\right)\right)}]}.
\end{equation}
Solving this for a given $A$ gives rise to the value $x_A$, which leads to the maximal value $j_{1}(x_A)$. Thus, with Eq.~(\ref{eq:J_j_x}) we find $t_{f}=2Aj_{1}(x_A) / J$, which for $F_T = 0.99$ leads to $t_p=2t_f \approx 12/J$.

\begin{figure}[h]
\includegraphics[width=\linewidth]{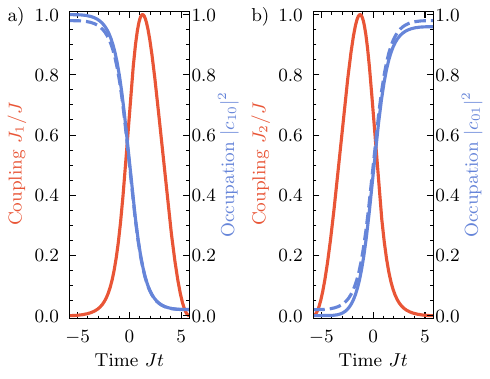}
\caption{\label{fig:prot_g_t}
In red: effective coupling a) $J_1$ and b) $J_2$ normalized by its maximal value $J$ as a function of time $t$ normalized by $J$.
In blue: simulated occupation (solid line) a) $|c_{10}(t)|^{2}$ and b) $|c_{01}(t)|^{2}$ and the ideal occupation as per Eq.~(\ref{eq:c_f}) (dashed line) a) $1-f(t)$ and b) $f(t)$.
In both figures we used $J t_f = 5.8$, $\alpha=0$ and $\beta = 1$.}
\end{figure}

\section{Properties of the magnon mode}\label{app:magnon}

In this Appendix, we review the properties of the chiral magnon modes considered in Sec.~\ref{sec:imp}, as derived in Ref.~\onlinecite{dolsSteadystateEntanglementSpin2026}.
Including exchange interaction, the magnon modes propagating along the $y$ axis obey following dispersion relation
\begin{equation}
\omega_{k}=\sqrt{\omega_{{\rm H}}'\left(\omega_{{\rm H}}'+\omega_{{\rm M}}\right)+\dfrac{\omega_{{\rm M}}^{2}}{2}\left(\frac{1}{2}-\frac{e^{-2\left|k\right|d}}{2}\right)},
\label{eq:omega_k}
\end{equation}
where $\omega_{{\rm H}}'=\omega_{{\rm H}}+\omega_{{\rm M}}D_{{\rm ex}}k^{2}$ and $\omega_{{\rm M}}=\mu_{0}\gamma_{s}M_{s}$.
The associated magnetic field fluctuations at the upper surface $x>d/2$ of the YIG stripe are given by
\begin{equation}
\delta\boldsymbol{H}_{k}(\boldsymbol{r})=\delta H_{k}(\boldsymbol{r})\times\begin{cases}
\boldsymbol{e}_{-}, & k>0,\\
\boldsymbol{e}_{+}, & k<0,
\end{cases}
\end{equation}
with $\delta H_{k}(\boldsymbol{r})=\sqrt{2}Cke^{-x|k|+iky}$ and
\begin{equation}
    C =iA\frac{(|k|-\chi_{\mathrm{a}}k)\sinh(kd)+(1+\chi_{\mathrm{d}})k\cosh(kd)}{\left(|k|-\chi_{\mathrm{a}}k\right)\cosh\left(\frac{kd}{2}\right)+(1+\chi_{\mathrm{d}})k\sinh\left(\frac{kd}{2}\right)}e^{\frac{1}{2}|k|d}.
\end{equation}
Here, $\chi_{{\rm d}}$ and $\chi_{{\rm a}}$ are the diagonal and antidiagonal components of the Polder susceptibility tensor~\cite{stancilSpinWavesTheory2009}, which are defined by
\begin{eqnarray}
\chi_{{\rm d}} & = & \frac{\omega_{{\rm M}}\left(\omega_{{\rm H}}+\omega_{{\rm M}}k^{2}D_{{\rm ex}}\right)}{\left(\omega_{{\rm H}}+\omega_{{\rm M}}k^{2}D_{{\rm ex}}\right)^{2}-\omega_{k}^{2}},\\
\chi_{{\rm a}} & = & \frac{\omega_{k}\omega_{{\rm M}}}{\left(\omega_{{\rm H}}+\omega_{{\rm M}}k^{2}D_{{\rm ex}}\right)^{2}-\omega_{k}^{2}}.
\end{eqnarray}
\begin{widetext}
Furthermore,
\begin{equation}
A=-\operatorname{sgn}(k)\sqrt{\frac{\gamma\hbar M_{\mathrm{s}}}{2L_yL_zk\mathfrak{a}_{1}\mathfrak{a}_{2}\sinh(kd)}}\left\{ \operatorname{sgn}(k)\left[\chi_{\mathrm{a}}+(1+\chi_{\mathrm{d}})\tanh\left(\frac{kd}{2}\right)\right]+1\right\},
\end{equation}
with 
\begin{align*}
\mathfrak{a}_{1} & =\left[\chi_{\mathrm{a}}-(\chi_{\mathrm{d}}^{2}+\chi_{\mathrm{d}}-\chi_{\mathrm{a}}^{2})\operatorname{sgn}(k)\right]\tanh\left(\frac{kd}{2}\right)+\chi_{\mathrm{a}}\operatorname{sgn}(k)-\chi_{\mathrm{d}},\\
\mathfrak{a}_{2} & =\left[\chi_{\mathrm{a}}\operatorname{sgn}(k)-\chi_{\mathrm{d}}\right]\tanh\left(\frac{kd}{2}\right)-(\chi_{\mathrm{d}}^{2}+\chi_{\mathrm{d}}-\chi_{\mathrm{a}}^{2})\operatorname{sgn}(k)+\chi_{\mathrm{a}}.
\end{align*}
\end{widetext}
The magnetic field fluctuations $\delta\boldsymbol{H}_{k}(\boldsymbol{r})$ are used to compute the coupling given in Eq.~(\ref{eq:H_Q_M}).
\bibliography{refs.bib}

@article{abobeihOnesecondCoherenceSingle2018,
  title = {One-Second Coherence for a Single Electron Spin Coupled to a Multi-Qubit Nuclear-Spin Environment},
  author = {Abobeih, M. H. and Cramer, J. and Bakker, M. A. and Kalb, N. and Markham, M. and Twitchen, D. J. and Taminiau, T. H.},
  year = 2018,
  month = jun,
  journal = {Nature Communications},
  volume = {9},
  number = {1},
  pages = {2552},
  publisher = {Nature Publishing Group},
  issn = {2041-1723},
  doi = {10.1038/s41467-018-04916-z},
  copyright = {2018 The Author(s)},
  langid = {english}
}

@article{anCoherentLongrangeTransfer2020,
  title = {Coherent Long-Range Transfer of Angular Momentum between Magnon {{Kittel}} Modes by Phonons},
  author = {An, K. and Litvinenko, A. N. and Kohno, R. and Fuad, A. A. and Naletov, V. V. and Vila, L. and Ebels, U. and {de Loubens}, G. and Hurdequint, H. and Beaulieu, N. and Ben Youssef, J. and Vukadinovic, N. and Bauer, G. E. W. and Slavin, A. N. and Tiberkevich, V. S. and Klein, O.},
  year = 2020,
  month = feb,
  journal = {Physical Review B},
  volume = {101},
  number = {6},
  pages = {060407},
  doi = {10.1103/PhysRevB.101.060407}
}

@article{arrazolaEngineeringProtectedCavityQED2025a,
  title = {Engineering Protected Cavity-{{QED}} Interactions through Pulsed Dynamical Decoupling},
  author = {Arrazola, I. and Bertet, P. and Chu, Y. and Rabl, P.},
  year = 2025,
  month = nov,
  journal = {npj Quantum Information},
  volume = {11},
  number = {1},
  pages = {197},
  publisher = {Nature Publishing Group},
  issn = {2056-6387},
  doi = {10.1038/s41534-025-01143-5},
  copyright = {2025 The Author(s)},
  langid = {english}
}

@article{awschalomQuantumTechnologiesOptically2018a,
  title = {Quantum Technologies with Optically Interfaced Solid-State Spins},
  author = {Awschalom, David D. and Hanson, Ronald and Wrachtrup, J{\"o}rg and Zhou, Brian B.},
  year = 2018,
  month = sep,
  journal = {Nature Photonics},
  volume = {12},
  number = {9},
  pages = {516--527},
  publisher = {Nature Publishing Group},
  issn = {1749-4893},
  doi = {10.1038/s41566-018-0232-2},
  copyright = {2018 Springer Nature Limited},
  langid = {english}
}

@article{bar-gillSolidstateElectronicSpin2013,
  title = {Solid-State Electronic Spin Coherence Time Approaching One Second},
  author = {{Bar-Gill}, N. and Pham, L. M. and Jarmola, A. and Budker, D. and Walsworth, R. L.},
  year = 2013,
  month = apr,
  journal = {Nature Communications},
  volume = {4},
  number = {1},
  pages = {1743},
  issn = {2041-1723},
  doi = {10.1038/ncomms2771},
  copyright = {2013 Springer Nature Limited}
}

@article{bejaranoParametricMagnonTransduction2024,
  title = {Parametric Magnon Transduction to Spin Qubits},
  author = {Bejarano, Mauricio and Goncalves, Francisco J. T. and Hache, Toni and Hollenbach, Michael and Heins, Christopher and Hula, Tobias and K{\"o}rber, Lukas and Heinze, Jakob and Berenc{\'e}n, Yonder and Helm, Manfred and Fassbender, J{\"u}rgen and Astakhov, Georgy V. and Schultheiss, Helmut},
  year = 2024,
  month = mar,
  journal = {Science Advances},
  volume = {10},
  number = {12},
  pages = {eadi2042},
  doi = {10.1126/sciadv.adi2042}
}

@article{belmeguenaiInterfacialDzyaloshinskiiMoriyaInteraction2015,
  title = {Interfacial {{Dzyaloshinskii-Moriya}} Interaction in Perpendicularly Magnetized {{Pt/Co/AlO{\textsubscript{x}}}} Ultrathin Films Measured by {{Brillouin}} Light Spectroscopy},
  author = {Belmeguenai, Mohamed and Adam, Jean-Paul and Roussign{\'e}, Yves and Eimer, Sylvain and Devolder, Thibaut and Kim, Joo-Von and Cherif, Salim Mourad and Stashkevich, Andrey and Thiaville, Andr{\'e}},
  year = 2015,
  month = may,
  journal = {Physical Review B},
  volume = {91},
  number = {18},
  pages = {180405},
  doi = {10.1103/PhysRevB.91.180405}
}

@article{bensmannDispersiontunableLowlossImplanted2025,
  title = {Dispersion-Tunable Low-Loss Implanted Spin-Wave Waveguides for Large Magnonic Networks},
  author = {Bensmann, Jannis and Schmidt, Robert and Nikolaev, Kirill O. and Raskhodchikov, Dimitri and Choudhary, Shraddha and Bhardwaj, Richa and Taheriniya, Shabnam and Varri, Akhil and Niehues, Sven and El Kadri, Ahmad and Kern, Johannes and Pernice, Wolfram H. P. and Demokritov, Sergej O. and Demidov, Vladislav E. and {Michaelis de Vasconcellos}, Steffen and Bratschitsch, Rudolf},
  year = 2025,
  month = dec,
  journal = {Nature Materials},
  volume = {24},
  number = {12},
  pages = {1920--1926},
  publisher = {Nature Publishing Group},
  issn = {1476-4660},
  doi = {10.1038/s41563-025-02282-y},
  copyright = {2025 The Author(s)},
  langid = {english}
}

@article{bertelliMagneticResonanceImaging2020,
  title = {Magnetic Resonance Imaging of Spin-Wave Transport and Interference in a Magnetic Insulator},
  author = {Bertelli, Iacopo and Carmiggelt, Joris J. and Yu, Tao and Simon, Brecht G. and Pothoven, Coosje C. and Bauer, Gerrit E. W. and Blanter, Yaroslav M. and Aarts, Jan and {van der Sar}, Toeno},
  year = 2020,
  month = nov,
  journal = {Science Advances},
  volume = {6},
  number = {46},
  pages = {eabd3556},
  doi = {10.1126/sciadv.abd3556}
}

@article{bittencourtOptomagnonicsDispersiveMedia2022,
  title = {Optomagnonics in {{Dispersive Media}}: {{Magnon-Photon Coupling Enhancement}} at the {{Epsilon-near-Zero Frequency}}},
  shorttitle = {Optomagnonics in {{Dispersive Media}}},
  author = {Bittencourt, V. A. S. V. and Liberal, I. and Viola Kusminskiy, S.},
  year = 2022,
  month = may,
  journal = {Physical Review Letters},
  volume = {128},
  number = {18},
  pages = {183603},
  publisher = {American Physical Society},
  doi = {10.1103/PhysRevLett.128.183603}
}

@book{breuerTheoryOpenQuantum2007,
  title = {The {{Theory}} of {{Open Quantum Systems}}},
  author = {Breuer, Heinz-Peter and Petruccione, Francesco},
  year = 2007,
  month = jan,
  edition = {1},
  publisher = {Oxford University PressOxford},
  doi = {10.1093/acprof:oso/9780199213900.001.0001},
  isbn = {978-0-19-921390-0 978-0-19-170634-9}
}

@article{bruhlmannClassicalQuantumTheory2026,
  title = {Classical and Quantum Theory of Magnonic and Magnetoelastic Nonlinear Dynamics in Continuum Geometries},
  author = {Br{\"u}hlmann, Marco and Hwang, Yunyoung and Puebla, Jorge and {Gonzalez-Ballestero}, Carlos},
  year = 2026,
  month = jan,
  journal = {Physical Review B},
  volume = {113},
  number = {1},
  pages = {014430},
  publisher = {American Physical Society},
  doi = {10.1103/d12l-krcy}
}

@article{caiDynamicEntanglementOscillating2010,
  title = {Dynamic Entanglement in Oscillating Molecules and Potential Biological Implications},
  author = {Cai, Jianming and Popescu, Sandu and Briegel, Hans J.},
  year = 2010,
  month = aug,
  journal = {Physical Review E},
  volume = {82},
  number = {2},
  pages = {021921},
  publisher = {American Physical Society},
  doi = {10.1103/PhysRevE.82.021921}
}

@article{caiRobustDynamicalDecoupling2012,
  title = {Robust Dynamical Decoupling with Concatenated Continuous Driving},
  author = {Cai, J-M and Naydenov, B and Pfeiffer, R and McGuinness, L P and Jahnke, K D and Jelezko, F and Plenio, M B and Retzker, A},
  year = 2012,
  month = nov,
  journal = {New Journal of Physics},
  volume = {14},
  number = {11},
  pages = {113023},
  publisher = {IOP Publishing},
  issn = {1367-2630},
  doi = {10.1088/1367-2630/14/11/113023},
  langid = {english}
}

@article{casolaProbingCondensedMatter2018,
  title = {Probing Condensed Matter Physics with Magnetometry Based on Nitrogen-Vacancy Centres in Diamond},
  author = {Casola, Francesco and {van der Sar}, Toeno and Yacoby, Amir},
  year = 2018,
  month = jan,
  journal = {Nature Reviews Materials},
  volume = {3},
  number = {1},
  pages = {1--13},
  issn = {2058-8437},
  doi = {10.1038/natrevmats.2017.88},
  copyright = {2018 Macmillan Publishers Limited}
}

@article{cherepanovSagaYIGSpectra1993,
  title = {The Saga of {{YIG}}: {{Spectra}}, Thermodynamics, Interaction and Relaxation of Magnons in a Complex Magnet},
  shorttitle = {The Saga of {{YIG}}},
  author = {Cherepanov, Vladimir and Kolokolov, Igor and L'vov, Victor},
  year = 1993,
  month = jul,
  journal = {Physics Reports},
  volume = {229},
  number = {3},
  pages = {81--144},
  issn = {0370-1573},
  doi = {10.1016/0370-1573(93)90107-O}
}

@article{chumakAdvancesMagneticsRoadmap2022,
  title = {Advances in {{Magnetics Roadmap}} on {{Spin-Wave Computing}}},
  author = {Chumak, A. V. and Kabos, P. and Wu, M. and Abert, C. and Adelmann, C. and Adeyeye, A. O. and {\AA}kerman, J. and Aliev, F. G. and Anane, A. and Awad, A. and Back, C. H. and Barman, A. and Bauer, G. E. W. and Becherer, M. and Beginin, E. N. and Bittencourt, V. A. S. V. and Blanter, Y. M. and Bortolotti, P. and Boventer, I. and Bozhko, D. A. and Bunyaev, S. A. and Carmiggelt, J. J. and Cheenikundil, R. R. and Ciubotaru, F. and Cotofana, S. and Csaba, G. and Dobrovolskiy, O. V. and Dubs, C. and Elyasi, M. and Fripp, K. G. and Fulara, H. and Golovchanskiy, I. A. and {Gonzalez-Ballestero}, C. and Graczyk, P. and Grundler, D. and Gruszecki, P. and Gubbiotti, G. and Guslienko, K. and Haldar, A. and Hamdioui, S. and Hertel, R. and Hillebrands, B. and Hioki, T. and Houshang, A. and Hu, C.-M. and Huebl, H. and Huth, M. and Iacocca, E. and Jungfleisch, M. B. and Kakazei, G. N. and Khitun, A. and Khymyn, R. and Kikkawa, T. and Kl{\"a}ui, M. and Klein, O. and K{\l}os, J. W. and Knauer, S. and Koraltan, S. and Kostylev, M. and Krawczyk, M. and Krivorotov, I. N. and Kruglyak, V. V. and {Lachance-Quirion}, D. and Ladak, S. and Lebrun, R. and Li, Y. and Lindner, M. and Mac{\^e}do, R. and Mayr, S. and Melkov, G. A. and Mieszczak, S. and Nakamura, Y. and Nembach, H. T. and Nikitin, A. A. and Nikitov, S. A. and Novosad, V. and Ot{\'a}lora, J. A. and Otani, Y. and Papp, A. and Pigeau, B. and Pirro, P. and Porod, W. and Porrati, F. and Qin, H. and Rana, B. and Reimann, T. and Riente, F. and {Romero-Isart}, O. and Ross, A. and Sadovnikov, A. V. and Safin, A. R. and Saitoh, E. and Schmidt, G. and Schultheiss, H. and Schultheiss, K. and Serga, A. A. and Sharma, S. and Shaw, J. M. and Suess, D. and Surzhenko, O. and Szulc, K. and Taniguchi, T. and Urb{\'a}nek, M. and Usami, K. and Ustinov, A. B. and {van der Sar}, T. and {van Dijken}, S. and Vasyuchka, V. I. and Verba, R. and Kusminskiy, S. Viola and Wang, Q. and Weides, M. and Weiler, M. and Wintz, S. and Wolski, S. P. and Zhang, X.},
  year = 2022,
  month = jun,
  journal = {IEEE Transactions on Magnetics},
  volume = {58},
  number = {6},
  pages = {1--72},
  issn = {1941-0069},
  doi = {10.1109/TMAG.2022.3149664}
}

@article{chumakMagnonSpintronics2015,
  title = {Magnon Spintronics},
  author = {Chumak, A. V. and Vasyuchka, V. I. and Serga, A. A. and Hillebrands, B.},
  year = 2015,
  month = jun,
  journal = {Nature Physics},
  volume = {11},
  number = {6},
  pages = {453--461},
  publisher = {Nature Publishing Group},
  issn = {1745-2481},
  doi = {10.1038/nphys3347},
  copyright = {2014 Springer Nature Limited},
  langid = {english}
}

@article{ciracQuantumStateTransfer1997,
  title = {Quantum {{State Transfer}} and {{Entanglement Distribution}} among {{Distant Nodes}} in a {{Quantum Network}}},
  author = {Cirac, J. I. and Zoller, P. and Kimble, H. J. and Mabuchi, H.},
  year = 1997,
  month = apr,
  journal = {Physical Review Letters},
  volume = {78},
  number = {16},
  pages = {3221--3224},
  publisher = {American Physical Society},
  doi = {10.1103/PhysRevLett.78.3221}
}

@article{cortes-ortunoInfluenceDzyaloshinskiiMoriya2013,
  title = {Influence of the {{Dzyaloshinskii}}--{{Moriya}} Interaction on the Spin-Wave Spectra of Thin Films},
  author = {{Cort{\'e}s-Ortu{\~n}o}, D and Landeros, P},
  year = 2013,
  month = mar,
  journal = {Journal of Physics: Condensed Matter},
  volume = {25},
  number = {15},
  pages = {156001},
  issn = {0953-8984},
  doi = {10.1088/0953-8984/25/15/156001}
}

@article{damonMagnetostaticModesFerromagnet1961,
  title = {Magnetostatic Modes of a Ferromagnet Slab},
  author = {Damon, R. W. and Eshbach, J. R.},
  year = 1961,
  month = may,
  journal = {Journal of Physics and Chemistry of Solids},
  volume = {19},
  number = {3},
  pages = {308--320},
  issn = {0022-3697},
  doi = {10.1016/0022-3697(61)90041-5}
}

@misc{deySensingMagnonicQuantum2025,
  title = {Sensing Magnonic Quantum Superpositions Using a Bosonic Mode as the Probe},
  author = {Dey, Bashab and Verma, Sonu and Weiler, Mathias and Kamra, Akashdeep},
  year = 2025,
  month = jul,
  number = {arXiv:2507.19066},
  eprint = {2507.19066},
  primaryclass = {cond-mat.mes-hall},
  publisher = {arXiv},
  doi = {10.48550/arXiv.2507.19066},
  archiveprefix = {arXiv}
}

@article{dohertyNitrogenvacancyColourCentre2013,
  title = {The Nitrogen-Vacancy Colour Centre in Diamond},
  author = {Doherty, Marcus W. and Manson, Neil B. and Delaney, Paul and Jelezko, Fedor and Wrachtrup, J{\"o}rg and Hollenberg, Lloyd C. L.},
  year = 2013,
  month = jul,
  journal = {Physics Reports},
  series = {The Nitrogen-Vacancy Colour Centre in Diamond},
  volume = {528},
  number = {1},
  pages = {1--45},
  issn = {0370-1573},
  doi = {10.1016/j.physrep.2013.02.001}
}

@article{dolsMagnonmediatedQuantumGates2024a,
  title = {Magnon-Mediated Quantum Gates for Superconducting Qubits},
  author = {Dols, Martijn and Sharma, Sanchar and Bechara, Lenos and Blanter, Yaroslav M. and Kounalakis, Marios and Viola Kusminskiy, Silvia},
  year = 2024,
  month = sep,
  journal = {Physical Review B},
  volume = {110},
  number = {10},
  pages = {104416},
  doi = {10.1103/PhysRevB.110.104416}
}

@article{dolsSteadystateEntanglementSpin2026,
  title = {Steady-State Entanglement of Spin Qubits Mediated by Nonreciprocal and Chiral Magnons},
  author = {Dols, Martijn and Cherkasskii, Mikhail and Bittencourt, Victor A. S. V. and {Gonzalez-Ballestero}, Carlos and Dasari, Durga B. R. and Viola Kusminskiy, Silvia},
  year = 2026,
  month = jun,
  journal = {Physical Review Research},
  volume = {8},
  number = {2},
  pages = {023310},
  publisher = {American Physical Society},
  doi = {10.1103/w358-y2z9}
}

@article{flebus2024MagnonicsRoadmap2024,
  title = {The 2024 Magnonics Roadmap},
  author = {Flebus, Benedetta and Grundler, Dirk and Rana, Bivas and Otani, YoshiChika and Barsukov, Igor and Barman, Anjan and Gubbiotti, Gianluca and Landeros, Pedro and Akerman, Johan and Ebels, Ursula and Pirro, Philipp and Demidov, Vladislav E and Schultheiss, Katrin and Csaba, Gyorgy and Wang, Qi and Ciubotaru, Florin and Nikonov, Dmitri E and Che, Ping and Hertel, Riccardo and Ono, Teruo and Afanasiev, Dmytro and Mentink, Johan and Rasing, Theo and Hillebrands, Burkard and Kusminskiy, Silvia Viola and Zhang, Wei and Du, Chunhui Rita and Finco, Aurore and {van der Sar}, Toeno and Luo, Yunqiu Kelly and Shiota, Yoichi and Sklenar, Joseph and Yu, Tao and Rao, Jinwei},
  year = 2024,
  month = jun,
  journal = {Journal of Physics: Condensed Matter},
  volume = {36},
  number = {36},
  pages = {363501},
  issn = {0953-8984},
  doi = {10.1088/1361-648X/ad399c}
}

@article{flebusEntanglingDistantSpin2019,
  title = {Entangling Distant Spin Qubits via a Magnetic Domain Wall},
  author = {Flebus, B. and Tserkovnyak, Y.},
  year = 2019,
  month = apr,
  journal = {Physical Review B},
  volume = {99},
  number = {14},
  pages = {140403},
  doi = {10.1103/PhysRevB.99.140403}
}

@article{frydrychSelectiveDynamicalDecoupling2014,
  title = {Selective Dynamical Decoupling for Quantum State Transfer},
  author = {Frydrych, H and Hoskovec, A and Jex, I and Alber, G},
  year = 2014,
  month = dec,
  journal = {Journal of Physics B: Atomic, Molecular and Optical Physics},
  volume = {48},
  number = {2},
  pages = {025501},
  publisher = {IOP Publishing},
  issn = {0953-4075},
  doi = {10.1088/0953-4075/48/2/025501},
  langid = {english}
}

@article{fukamiMagnonmediatedQubitCoupling2024,
  title = {Magnon-Mediated Qubit Coupling Determined via Dissipation Measurements},
  author = {Fukami, Masaya and Marcks, Jonathan C. and Candido, Denis R. and Weiss, Leah R. and Soloway, Benjamin and Sullivan, Sean E. and Delegan, Nazar and Heremans, F. Joseph and Flatt{\'e}, Michael E. and Awschalom, David D.},
  year = 2024,
  month = jan,
  journal = {Proceedings of the National Academy of Sciences},
  volume = {121},
  number = {2},
  pages = {e2313754120},
  doi = {10.1073/pnas.2313754120}
}

@article{fukamiOpportunitiesLongRangeMagnonMediated2021,
  title = {Opportunities for {{Long-Range Magnon-Mediated Entanglement}} of {{Spin Qubits}} via {{On-}} and {{Off-Resonant Coupling}}},
  author = {Fukami, Masaya and Candido, Denis R. and Awschalom, David D. and Flatt{\'e}, Michael E.},
  year = 2021,
  month = oct,
  journal = {PRX Quantum},
  volume = {2},
  number = {4},
  pages = {040314},
  doi = {10.1103/PRXQuantum.2.040314}
}

@article{gallardoCoherentMagnonsGiant2024,
  title = {Coherent {{Magnons}} with {{Giant Nonreciprocity}} at {{Nanoscale Wavelengths}}},
  author = {Gallardo, Rodolfo and Weigand, Markus and Schultheiss, Katrin and Kakay, Attila and Mattheis, Roland and Raabe, J{\"o}rg and Sch{\"u}tz, Gisela and Deac, Alina and Lindner, J{\"u}rgen and Wintz, Sebastian},
  year = 2024,
  month = feb,
  journal = {ACS Nano},
  volume = {18},
  number = {7},
  pages = {5249--5257},
  publisher = {American Chemical Society},
  issn = {1936-0851},
  doi = {10.1021/acsnano.3c08390}
}

@article{gallardoFlatBandsIndirect2019,
  title = {Flat {{Bands}}, {{Indirect Gaps}}, and {{Unconventional Spin-Wave Behavior Induced}} by a {{Periodic Dzyaloshinskii-Moriya Interaction}}},
  author = {Gallardo, R. A. and {Cort{\'e}s-Ortu{\~n}o}, D. and Schneider, T. and {Rold{\'a}n-Molina}, A. and Ma, Fusheng and Troncoso, R. E. and Lenz, K. and Fangohr, H. and Lindner, J. and Landeros, P.},
  year = 2019,
  month = feb,
  journal = {Physical Review Letters},
  volume = {122},
  number = {6},
  pages = {067204},
  doi = {10.1103/PhysRevLett.122.067204}
}

@article{gallardoReconfigurableSpinWaveNonreciprocity2019,
  title = {Reconfigurable {{Spin-Wave Nonreciprocity Induced}} by {{Dipolar Interaction}} in a {{Coupled Ferromagnetic Bilayer}}},
  author = {Gallardo, R.A. and Schneider, T. and Chaurasiya, A.K. and Oelschl{\"a}gel, A. and Arekapudi, S.S.P.K. and {Rold{\'a}n-Molina}, A. and H{\"u}bner, R. and Lenz, K. and Barman, A. and Fassbender, J. and Lindner, J. and Hellwig, O. and Landeros, P.},
  year = 2019,
  month = sep,
  journal = {Physical Review Applied},
  volume = {12},
  number = {3},
  pages = {034012},
  doi = {10.1103/PhysRevApplied.12.034012}
}

@article{gallardoSpinwaveNonreciprocityMagnetizationgraded2019,
  title = {Spin-Wave Non-Reciprocity in Magnetization-Graded Ferromagnetic Films},
  author = {Gallardo, R A and {Alvarado-Seguel}, P and Schneider, T and {Gonzalez-Fuentes}, C and {Rold{\'a}n-Molina}, A and Lenz, K and Lindner, J and Landeros, P},
  year = 2019,
  month = mar,
  journal = {New Journal of Physics},
  volume = {21},
  number = {3},
  pages = {033026},
  issn = {1367-2630},
  doi = {10.1088/1367-2630/ab0449}
}

@article{Gonzalez-Ballestero_2013,
  title = {Non-{{Markovian}} Effects in Waveguide-Mediated Entanglement},
  author = {{Gonzalez-Ballestero}, C and {Garc{\'i}a-Vidal}, F J and Moreno, Esteban},
  year = 2013,
  month = jul,
  journal = {New Journal of Physics},
  volume = {15},
  number = {7},
  pages = {073015},
  publisher = {IOP Publishing},
  doi = {10.1088/1367-2630/15/7/073015}
}

@article{gonzalez-ballesteroQuantumAcoustomechanicsMicromagnet2020,
  title = {Quantum {{Acoustomechanics}} with a {{Micromagnet}}},
  author = {{Gonzalez-Ballestero}, Carlos and Gieseler, Jan and {Romero-Isart}, Oriol},
  year = 2020,
  month = mar,
  journal = {Physical Review Letters},
  volume = {124},
  number = {9},
  pages = {093602},
  publisher = {American Physical Society},
  doi = {10.1103/PhysRevLett.124.093602}
}

@article{gonzalez-ballesteroQuantumInterfaceSpin2022,
  title = {Towards a Quantum Interface between Spin Waves and Paramagnetic Spin Baths},
  author = {{Gonzalez-Ballestero}, C. and Van Der Sar, Toeno and {Romero-Isart}, O.},
  year = 2022,
  month = feb,
  journal = {Physical Review B},
  volume = {105},
  number = {7},
  pages = {075410},
  issn = {2469-9950, 2469-9969},
  doi = {10.1103/PhysRevB.105.075410}
}

@article{gonzalez-ballesteroTheoryQuantumAcoustomagnonics2020,
  title = {Theory of Quantum Acoustomagnonics and Acoustomechanics with a Micromagnet},
  author = {{Gonzalez-Ballestero}, C. and H{\"u}mmer, D. and Gieseler, J. and {Romero-Isart}, O.},
  year = 2020,
  month = mar,
  journal = {Physical Review B},
  volume = {101},
  number = {12},
  pages = {125404},
  publisher = {American Physical Society},
  doi = {10.1103/PhysRevB.101.125404}
}

@article{gonzalez-tudelaLightMatterInteractions2024,
  title = {Light--Matter Interactions in Quantum Nanophotonic Devices},
  author = {{Gonz{\'a}lez-Tudela}, Alejandro and Reiserer, Andreas and {Garc{\'i}a-Ripoll}, Juan Jos{\'e} and {Garc{\'i}a-Vidal}, Francisco J.},
  year = 2024,
  month = mar,
  journal = {Nature Reviews Physics},
  volume = {6},
  number = {3},
  pages = {166--179},
  publisher = {Nature Publishing Group},
  issn = {2522-5820},
  doi = {10.1038/s42254-023-00681-1},
  copyright = {2024 Springer Nature Limited},
  langid = {english}
}

@article{goryachevHighCooperativityCavityQED2014,
  title = {High-{{Cooperativity Cavity QED}} with {{Magnons}} at {{Microwave Frequencies}}},
  author = {Goryachev, Maxim and Farr, Warrick G. and Creedon, Daniel L. and Fan, Yaohui and Kostylev, Mikhail and Tobar, Michael E.},
  year = 2014,
  month = nov,
  journal = {Physical Review Applied},
  volume = {2},
  number = {5},
  pages = {054002},
  doi = {10.1103/PhysRevApplied.2.054002}
}

@article{hacheNanoscaleMappingMagnetic2025,
  title = {Nanoscale {{Mapping}} of {{Magnetic Auto-Oscillations}} with a {{Single Spin Sensor}}},
  author = {Hache, Toni and Anshu, Anshu and Shalomayeva, Tetyana and Richter, Gunther and St{\"o}hr, Rainer and Kern, Klaus and Wrachtrup, J{\"o}rg and Singha, Aparajita},
  year = 2025,
  month = feb,
  journal = {Nano Letters},
  volume = {25},
  number = {5},
  pages = {1917--1924},
  publisher = {American Chemical Society},
  issn = {1530-6984},
  doi = {10.1021/acs.nanolett.4c05531}
}

@article{haighSelectionRulesCavityenhanced2018,
  title = {Selection Rules for Cavity-Enhanced {{Brillouin}} Light Scattering from Magnetostatic Modes},
  author = {Haigh, J. A. and Lambert, N. J. and Sharma, S. and Blanter, Y. M. and Bauer, G. E. W. and Ramsay, A. J.},
  year = 2018,
  month = jun,
  journal = {Physical Review B},
  volume = {97},
  number = {21},
  pages = {214423},
  doi = {10.1103/PhysRevB.97.214423}
}

@article{heinsNonreciprocalSpinwaveDispersion2025,
  title = {Nonreciprocal Spin-Wave Dispersion in Magnetic Bilayers},
  author = {Heins, Christopher and Iurchuk, Vadym and Gladii, Olga and K{\"o}rber, Lukas and K{\'a}kay, Attila and Fassbender, J{\"u}rgen and Schultheiss, Katrin and Schultheiss, Helmut},
  year = 2025,
  month = apr,
  journal = {Physical Review B},
  volume = {111},
  number = {13},
  pages = {134434},
  doi = {10.1103/PhysRevB.111.134434}
}

@article{hetenyiLongdistanceCouplingSpin2022,
  title = {Long-Distance Coupling of Spin Qubits via Topological Magnons},
  author = {Het{\'e}nyi, Bence and Mook, Alexander and Klinovaja, Jelena and Loss, Daniel},
  year = 2022,
  month = dec,
  journal = {Physical Review B},
  volume = {106},
  number = {23},
  pages = {235409},
  doi = {10.1103/PhysRevB.106.235409}
}

@article{hillebrandsSpinwaveCalculationsMultilayered1990,
  title = {Spin-Wave Calculations for Multilayered Structures},
  author = {Hillebrands, Burkard},
  year = 1990,
  month = jan,
  journal = {Physical Review B},
  volume = {41},
  number = {1},
  pages = {530--540},
  doi = {10.1103/PhysRevB.41.530}
}

@article{hueblHighCooperativityCoupled2013,
  title = {High {{Cooperativity}} in {{Coupled Microwave Resonator Ferrimagnetic Insulator Hybrids}}},
  author = {Huebl, Hans and Zollitsch, Christoph W. and Lotze, Johannes and Hocke, Fredrik and Greifenstein, Moritz and Marx, Achim and Gross, Rudolf and Goennenwein, Sebastian T. B.},
  year = 2013,
  month = sep,
  journal = {Physical Review Letters},
  volume = {111},
  number = {12},
  pages = {127003},
  doi = {10.1103/PhysRevLett.111.127003}
}

@article{hwangHarmonicSubharmonicMagnon2026,
  title = {Harmonic and Subharmonic Magnon Generation in a Surface-Acoustic-Wave Resonator},
  author = {Hwang, Yunyoung and Liao, Liyang and Puebla, Jorge and Br{\"u}hlmann, Marco and {Gonzalez-Ballestero}, Carlos and Kondou, Kouta and Ogawa, Naoki and Maekawa, Sadamichi and Otani, Yoshichika},
  year = 2026,
  month = mar,
  journal = {Physical Review Applied},
  volume = {25},
  number = {3},
  pages = {034056},
  publisher = {American Physical Society},
  doi = {10.1103/mm21-ctsb}
}

@article{ishibashiSwitchableGiantNonreciprocal2020,
  title = {Switchable Giant Nonreciprocal Frequency Shift of Propagating Spin Waves in Synthetic Antiferromagnets},
  author = {Ishibashi, Mio and Shiota, Yoichi and Li, Tian and Funada, Shinsaku and Moriyama, Takahiro and Ono, Teruo},
  year = 2020,
  month = apr,
  journal = {Science Advances},
  volume = {6},
  number = {17},
  pages = {eaaz6931},
  doi = {10.1126/sciadv.aaz6931}
}

@article{karanikolasMagnonmediatedSpinEntanglement2022,
  title = {Magnon-Mediated Spin Entanglement in the Strong-Coupling Regime},
  author = {Karanikolas, Vasilios and Kuroda, Takashi and Inoue, Jun-ichi},
  year = 2022,
  month = dec,
  journal = {Physical Review Research},
  volume = {4},
  number = {4},
  pages = {043180},
  doi = {10.1103/PhysRevResearch.4.043180}
}

@article{korberCurvilinearSpinwaveDynamics2022,
  title = {Curvilinear Spin-Wave Dynamics beyond the Thin-Shell Approximation: {{Magnetic}} Nanotubes as a Case Study},
  shorttitle = {Curvilinear Spin-Wave Dynamics beyond the Thin-Shell Approximation},
  author = {K{\"o}rber, L. and Verba, R. and Ot{\'a}lora, Jorge A. and Kravchuk, V. and Lindner, J. and Fassbender, J. and K{\'a}kay, A.},
  year = 2022,
  month = jul,
  journal = {Physical Review B},
  volume = {106},
  number = {1},
  pages = {014405},
  doi = {10.1103/PhysRevB.106.014405}
}

@article{kounalakisAnalogQuantumControl2022,
  title = {Analog {{Quantum Control}} of {{Magnonic Cat States}} on a {{Chip}} by a {{Superconducting Qubit}}},
  author = {Kounalakis, Marios and Bauer, Gerrit E. W. and Blanter, Yaroslav M.},
  year = 2022,
  month = jul,
  journal = {Physical Review Letters},
  volume = {129},
  number = {3},
  pages = {037205},
  doi = {10.1103/PhysRevLett.129.037205}
}

@article{lachance-quirionEntanglementbasedSingleshotDetection2020,
  title = {Entanglement-Based Single-Shot Detection of a Single Magnon with a Superconducting Qubit},
  author = {{Lachance-Quirion}, Dany and Wolski, Samuel Piotr and Tabuchi, Yutaka and Kono, Shingo and Usami, Koji and Nakamura, Yasunobu},
  year = 2020,
  month = jan,
  journal = {Science},
  volume = {367},
  number = {6476},
  pages = {425--428},
  doi = {10.1126/science.aaz9236}
}

@article{liuOptomagnonicsMagneticSolids2016,
  title = {Optomagnonics in Magnetic Solids},
  author = {Liu, Tianyu and Zhang, Xufeng and Tang, Hong X. and Flatt{\'e}, Michael E.},
  year = 2016,
  month = aug,
  journal = {Physical Review B},
  volume = {94},
  number = {6},
  pages = {060405},
  doi = {10.1103/PhysRevB.94.060405}
}

@article{lodahlChiralQuantumOptics2017,
  title = {Chiral Quantum Optics},
  author = {Lodahl, Peter and Mahmoodian, Sahand and Stobbe, S{\o}ren and Rauschenbeutel, Arno and Schneeweiss, Philipp and Volz, J{\"u}rgen and Pichler, Hannes and Zoller, Peter},
  year = 2017,
  month = jan,
  journal = {Nature},
  volume = {541},
  number = {7638},
  pages = {473--480},
  issn = {1476-4687},
  doi = {10.1038/nature21037},
  copyright = {2017 Macmillan Publishers Limited, part of Springer Nature. All rights reserved.}
}

@article{maletinskyRobustScanningDiamond2012,
  title = {A Robust Scanning Diamond Sensor for Nanoscale Imaging with Single Nitrogen-Vacancy Centres},
  author = {Maletinsky, P. and Hong, S. and Grinolds, M. S. and Hausmann, B. and Lukin, M. D. and Walsworth, R. L. and Loncar, M. and Yacoby, A.},
  year = 2012,
  month = may,
  journal = {Nature Nanotechnology},
  volume = {7},
  number = {5},
  pages = {320--324},
  publisher = {Nature Publishing Group},
  issn = {1748-3395},
  doi = {10.1038/nnano.2012.50},
  copyright = {2012 Springer Nature Limited},
  langid = {english}
}

@article{mullerChiralPhononsPhononic2024,
  title = {Chiral Phonons and Phononic Birefringence in Ferromagnetic Metal--Bulk Acoustic Resonator Hybrids},
  author = {M{\"u}ller, M.},
  year = 2024,
  journal = {Physical Review B},
  volume = {109},
  number = {2},
  pages = {024430},
  doi = {10.1103/PhysRevB.109.024430}
}

@misc{mycroftQuantumStatePreparation2026,
  title = {Quantum {{State Preparation}} of {{Ferromagnetic Magnons}} by {{Parametric Driving}}},
  author = {Mycroft, Monika E. and Serha, Rostyslav O. and Chumak, Andrii V. and {Gonzalez-Ballestero}, Carlos},
  year = 2026,
  month = jan,
  number = {arXiv:2601.12833},
  eprint = {2601.12833},
  primaryclass = {cond-mat.mes-hall},
  publisher = {arXiv},
  doi = {10.48550/arXiv.2601.12833},
  archiveprefix = {arXiv}
}

@article{neumanNanomagnonicCavitiesStrong2020,
  title = {Nanomagnonic {{Cavities}} for {{Strong Spin-Magnon Coupling}} and {{Magnon-Mediated Spin-Spin Interactions}}},
  author = {Neuman, Tom{\'a}{\v s} and Wang, Derek S. and Narang, Prineha},
  year = 2020,
  month = dec,
  journal = {Physical Review Letters},
  volume = {125},
  number = {24},
  pages = {247702},
  doi = {10.1103/PhysRevLett.125.247702}
}

@article{osadaCavityOptomagnonicsSpinOrbit2016,
  title = {Cavity {{Optomagnonics}} with {{Spin-Orbit Coupled Photons}}},
  author = {Osada, A. and Hisatomi, R. and Noguchi, A. and Tabuchi, Y. and Yamazaki, R. and Usami, K. and Sadgrove, M. and Yalla, R. and Nomura, M. and Nakamura, Y.},
  year = 2016,
  month = jun,
  journal = {Physical Review Letters},
  volume = {116},
  number = {22},
  pages = {223601},
  doi = {10.1103/PhysRevLett.116.223601}
}

@article{otaloraCurvatureInducedAsymmetricSpinWave2016,
  title = {Curvature-{{Induced Asymmetric Spin-Wave Dispersion}}},
  author = {Ot{\'a}lora, Jorge A. and Yan, Ming and Schultheiss, Helmut and Hertel, Riccardo and K{\'a}kay, Attila},
  year = 2016,
  month = nov,
  journal = {Physical Review Letters},
  volume = {117},
  number = {22},
  pages = {227203},
  doi = {10.1103/PhysRevLett.117.227203}
}

@article{parekhPropagationCharacteristicsMagnetostatic1985,
  title = {Propagation Characteristics of Magnetostatic Waves},
  author = {Parekh, J. P. and Chang, K. W. and Tuan, H. S.},
  year = 1985,
  month = mar,
  journal = {Circuits, Systems and Signal Processing},
  volume = {4},
  number = {1},
  pages = {9--39},
  issn = {1531-5878},
  doi = {10.1007/BF01600071}
}

@article{pengCavityMagnonPolariton2025,
  title = {Cavity Magnon--Polariton Interface for Strong Spin--Spin Coupling},
  author = {Peng, Ma-Lei and Tian, Miao and Chen, Xue-Chun and Wang, Ming-Feng and Zhang, Guo-Qiang and Li, Hai-Chao and Xiong, Wei},
  year = 2025,
  month = mar,
  journal = {Optics Letters},
  volume = {50},
  number = {5},
  pages = {1516--1519},
  issn = {1539-4794},
  doi = {10.1364/OL.545688},
  copyright = {\copyright{} 2025 Optica Publishing Group}
}

@article{pichlerQuantumOpticsChiral2015,
  title = {Quantum Optics of Chiral Spin Networks},
  author = {Pichler, Hannes and Ramos, Tom{\'a}s and Daley, Andrew J. and Zoller, Peter},
  year = 2015,
  month = apr,
  journal = {Physical Review A},
  volume = {91},
  number = {4},
  pages = {042116},
  issn = {1050-2947, 1094-1622},
  doi = {10.1103/PhysRevA.91.042116},
  copyright = {http://link.aps.org/licenses/aps-default-license}
}

@article{pottsDynamicalBackactionMagnomechanics2021,
  title = {Dynamical {{Backaction Magnomechanics}}},
  author = {Potts, C. A. and Varga, E. and Bittencourt, V. A. S. V. and Kusminskiy, S. Viola and Davis, J. P.},
  year = 2021,
  month = sep,
  journal = {Physical Review X},
  volume = {11},
  number = {3},
  pages = {031053},
  doi = {10.1103/PhysRevX.11.031053}
}

@article{romlingSqueezingQuantumControl2025,
  title = {Squeezing and Quantum Control of the Antiferromagnetic Magnon Pseudospin},
  author = {R{\"o}mling, Anna-Luisa E. and Feist, Johannes and {Garc{\'i}a-Vidal}, Francisco J. and Kamra, Akashdeep},
  year = 2025,
  month = dec,
  journal = {Physical Review B},
  volume = {112},
  number = {21},
  pages = {214444},
  publisher = {American Physical Society},
  doi = {10.1103/qjm2-d19g}
}

@article{rondinMagnetometryNitrogenvacancyDefects2014,
  title = {Magnetometry with Nitrogen-Vacancy Defects in Diamond},
  author = {Rondin, L and Tetienne, J-P and Hingant, T and Roch, J-F and Maletinsky, P and Jacques, V},
  year = 2014,
  month = may,
  journal = {Reports on Progress in Physics},
  volume = {77},
  number = {5},
  pages = {056503},
  publisher = {IOP Publishing},
  issn = {0034-4885},
  doi = {10.1088/0034-4885/77/5/056503},
  langid = {english}
}

@article{rondinNanoscaleMagneticField2012,
  title = {Nanoscale Magnetic Field Mapping with a Single Spin Scanning Probe Magnetometer},
  author = {Rondin, L. and Tetienne, J.-P. and Spinicelli, P. and Dal Savio, C. and Karrai, K. and Dantelle, G. and Thiaville, A. and Rohart, S. and Roch, J.-F. and Jacques, V.},
  year = 2012,
  month = apr,
  journal = {Applied Physics Letters},
  volume = {100},
  number = {15},
  pages = {153118},
  issn = {0003-6951},
  doi = {10.1063/1.3703128}
}

@article{rousochatzakisMasterEquationsPulsed2005,
  title = {Master Equations for Pulsed Magnetic Fields: {{Application}} to Magnetic Molecules},
  shorttitle = {Master Equations for Pulsed Magnetic Fields},
  author = {Rousochatzakis, Ioannis and Luban, Marshall},
  year = 2005,
  month = oct,
  journal = {Physical Review B},
  volume = {72},
  number = {13},
  pages = {134424},
  publisher = {American Physical Society},
  doi = {10.1103/PhysRevB.72.134424}
}

@article{schlitzMagnetizationDynamicsAffected2022,
  title = {Magnetization Dynamics Affected by Phonon Pumping},
  author = {Schlitz, Richard and Siegl, Luise and Sato, Takuma and Yu, Weichao and Bauer, Gerrit E. W. and Huebl, Hans and Goennenwein, Sebastian T. B.},
  year = 2022,
  month = jul,
  journal = {Physical Review B},
  volume = {106},
  number = {1},
  pages = {014407},
  doi = {10.1103/PhysRevB.106.014407}
}

@misc{serhaUltralonglivingMagnonsQuantum2025,
  title = {Ultra-Long-Living Magnons in the Quantum Limit},
  author = {Serha, Rostyslav O. and McAllister, Kaitlin H. and Majcen, Fabian and Knauer, Sebastian and Reimann, Timmy and Dubs, Carsten and Melkov, Gennadii A. and Serga, Alexander A. and Tyberkevych, Vasyl S. and Chumak, Andrii V. and Bozhko, Dmytro A.},
  year = 2025,
  month = may,
  number = {arXiv:2505.22773},
  eprint = {2505.22773},
  publisher = {arXiv},
  doi = {10.48550/arXiv.2505.22773},
  archiveprefix = {arXiv}
}

@article{sharmaSpinCatStates2021,
  title = {Spin Cat States in Ferromagnetic Insulators},
  author = {Sharma, Sanchar and Bittencourt, Victor A. S. V. and Karenowska, Alexy D. and Kusminskiy, Silvia Viola},
  year = 2021,
  month = mar,
  journal = {Physical Review B},
  volume = {103},
  number = {10},
  pages = {L100403},
  doi = {10.1103/PhysRevB.103.L100403}
}

@article{slukaEmissionPropagation1D2019,
  title = {Emission and Propagation of {{1D}} and {{2D}} Spin Waves with Nanoscale Wavelengths in Anisotropic Spin Textures},
  author = {Sluka, Volker and Schneider, Tobias and Gallardo, Rodolfo A. and K{\'a}kay, Attila and Weigand, Markus and Warnatz, Tobias and Mattheis, Roland and {Rold{\'a}n-Molina}, Alejandro and Landeros, Pedro and Tiberkevich, Vasil and Slavin, Andrei and Sch{\"u}tz, Gisela and Erbe, Artur and Deac, Alina and Lindner, J{\"u}rgen and Raabe, J{\"o}rg and Fassbender, J{\"u}rgen and Wintz, Sebastian},
  year = 2019,
  month = apr,
  journal = {Nature Nanotechnology},
  volume = {14},
  number = {4},
  pages = {328--333},
  issn = {1748-3395},
  doi = {10.1038/s41565-019-0383-4},
  copyright = {2019 The Author(s), under exclusive licence to Springer Nature Limited}
}

@article{soykalStrongFieldInteractions2010,
  title = {Strong {{Field Interactions}} between a {{Nanomagnet}} and a {{Photonic Cavity}}},
  author = {Soykal, {\"O}. O. and Flatt{\'e}, M. E.},
  year = {2010},
  month = feb,
  journal = {Physical Review Letters},
  volume = {104},
  number = {7},
  pages = {077202},
  doi = {10.1103/PhysRevLett.104.077202}
}

@book{stancilSpinWavesTheory2009,
  title = {Spin Waves: Theory and Applications},
  shorttitle = {Spin Waves},
  author = {Stancil, Daniel D. and Prabhakar, Anil},
  year = 2009,
  publisher = {Springer},
  address = {New York},
  isbn = {978-0-387-77864-8}
}

@article{stannigelDrivendissipativePreparationEntangled2012,
  title = {Driven-Dissipative Preparation of Entangled States in Cascaded Quantum-Optical Networks},
  author = {Stannigel, K and Rabl, P and Zoller, P},
  year = 2012,
  month = jun,
  journal = {New Journal of Physics},
  volume = {14},
  number = {6},
  pages = {063014},
  publisher = {IOP Publishing},
  issn = {1367-2630},
  doi = {10.1088/1367-2630/14/6/063014}
}

@article{stannigelOptomechanicalTransducersQuantuminformation2011a,
  title = {Optomechanical Transducers for Quantum-Information Processing},
  author = {Stannigel, K. and Rabl, P. and S{\o}rensen, A. S. and Lukin, M. D. and Zoller, P.},
  year = 2011,
  month = oct,
  journal = {Physical Review A},
  volume = {84},
  number = {4},
  pages = {042341},
  publisher = {American Physical Society},
  doi = {10.1103/PhysRevA.84.042341}
}

@article{suarez-foreroChiralQuantumOptics2025,
  title = {Chiral {{Quantum Optics}}: {{Recent Developments}} and {{Future Directions}}},
  shorttitle = {Chiral {{Quantum Optics}}},
  author = {{Su{\'a}rez-Forero}, D.G. and Jalali Mehrabad, M. and Vega, C. and {Gonz{\'a}lez-Tudela}, A. and Hafezi, M.},
  year = 2025,
  month = apr,
  journal = {PRX Quantum},
  volume = {6},
  number = {2},
  pages = {020101},
  doi = {10.1103/PRXQuantum.6.020101}
}

@article{sunMagneticDomainsDomain2021,
  title = {Magnetic Domains and Domain Wall Pinning in Atomically Thin {{CrBr3}} Revealed by Nanoscale Imaging},
  author = {Sun, Qi-Chao and Song, Tiancheng and Anderson, Eric and Brunner, Andreas and F{\"o}rster, Johannes and Shalomayeva, Tetyana and Taniguchi, Takashi and Watanabe, Kenji and Gr{\"a}fe, Joachim and St{\"o}hr, Rainer and Xu, Xiaodong and Wrachtrup, J{\"o}rg},
  year = 2021,
  month = mar,
  journal = {Nature Communications},
  volume = {12},
  number = {1},
  pages = {1989},
  publisher = {Nature Publishing Group},
  issn = {2041-1723},
  doi = {10.1038/s41467-021-22239-4},
  copyright = {2021 The Author(s)},
  langid = {english}
}

@article{tabuchiCoherentCouplingFerromagnetic2015,
  title = {Coherent Coupling between a Ferromagnetic Magnon and a Superconducting Qubit},
  author = {Tabuchi, Yutaka and Ishino, Seiichiro and Noguchi, Atsushi and Ishikawa, Toyofumi and Yamazaki, Rekishu and Usami, Koji and Nakamura, Yasunobu},
  year = 2015,
  month = jul,
  journal = {Science},
  volume = {349},
  number = {6246},
  pages = {405--408},
  doi = {10.1126/science.aaa3693}
}

@article{tabuchiHybridizingFerromagneticMagnons2014,
  title = {Hybridizing {{Ferromagnetic Magnons}} and {{Microwave Photons}} in the {{Quantum Limit}}},
  author = {Tabuchi, Yutaka and Ishino, Seiichiro and Ishikawa, Toyofumi and Yamazaki, Rekishu and Usami, Koji and Nakamura, Yasunobu},
  year = 2014,
  month = aug,
  journal = {Physical Review Letters},
  volume = {113},
  number = {8},
  pages = {083603},
  doi = {10.1103/PhysRevLett.113.083603}
}

@article{tacchiInterfacialDzyaloshinskiiMoriyaInteraction2017,
  title = {Interfacial {{Dzyaloshinskii-Moriya Interaction}} in {{Pt/CoFeB Films}}: {{Effect of the Heavy-Metal Thickness}}},
  shorttitle = {Interfacial {{Dzyaloshinskii-Moriya Interaction}} in \$\textbackslash mathrm\textbraceleft{{Pt}}\textbraceright/\textbackslash mathrm\textbraceleft{{CoFeB}}\textbraceright\$ {{Films}}},
  author = {Tacchi, S. and Troncoso, R. E. and Ahlberg, M. and Gubbiotti, G. and Madami, M. and {\AA}kerman, J. and Landeros, P.},
  year = 2017,
  month = apr,
  journal = {Physical Review Letters},
  volume = {118},
  number = {14},
  pages = {147201},
  doi = {10.1103/PhysRevLett.118.147201}
}

@article{thiancourtUnidirectionalSpinWaves2024,
  title = {Unidirectional Spin Waves Measured Using Propagating-Spin-Wave Spectroscopy},
  author = {Thiancourt, G.Y. and Ngom, S.M. and Bardou, N. and Devolder, T.},
  year = 2024,
  month = sep,
  journal = {Physical Review Applied},
  volume = {22},
  number = {3},
  pages = {034040},
  doi = {10.1103/PhysRevApplied.22.034040}
}

@article{trifunovicLongDistanceEntanglementSpin2013,
  title = {Long-{{Distance Entanglement}} of {{Spin Qubits}} via {{Ferromagnet}}},
  author = {Trifunovic, Luka and Pedrocchi, Fabio L. and Loss, Daniel},
  year = 2013,
  month = dec,
  journal = {Physical Review X},
  volume = {3},
  number = {4},
  pages = {041023},
  doi = {10.1103/PhysRevX.3.041023}
}

@article{udvardiChiralAsymmetrySpinWave2009,
  title = {Chiral {{Asymmetry}} of the {{Spin-Wave Spectra}} in {{Ultrathin Magnetic Films}}},
  author = {Udvardi, L. and Szunyogh, L.},
  year = 2009,
  month = may,
  journal = {Physical Review Letters},
  volume = {102},
  number = {20},
  pages = {207204},
  doi = {10.1103/PhysRevLett.102.207204}
}

@article{vermerschQuantumStateTransfer2017,
  title = {Quantum {{State Transfer}} via {{Noisy Photonic}} and {{Phononic Waveguides}}},
  author = {Vermersch, B. and Guimond, P.-O. and Pichler, H. and Zoller, P.},
  year = 2017,
  month = mar,
  journal = {Physical Review Letters},
  volume = {118},
  number = {13},
  pages = {133601},
  publisher = {American Physical Society},
  doi = {10.1103/PhysRevLett.118.133601}
}

@article{violaDynamicalDecouplingOpen1999,
  title = {Dynamical {{Decoupling}} of {{Open Quantum Systems}}},
  author = {Viola, Lorenza and Knill, Emanuel and Lloyd, Seth},
  year = 1999,
  month = mar,
  journal = {Physical Review Letters},
  volume = {82},
  number = {12},
  pages = {2417--2421},
  doi = {10.1103/PhysRevLett.82.2417}
}

@article{violakusminskiyCoupledSpinlightDynamics2016,
  title = {Coupled Spin-Light Dynamics in Cavity Optomagnonics},
  author = {Viola Kusminskiy, Silvia and Tang, Hong X. and Marquardt, Florian},
  year = 2016,
  month = sep,
  journal = {Physical Review A},
  volume = {94},
  number = {3},
  pages = {033821},
  doi = {10.1103/PhysRevA.94.033821}
}

@article{weberQuantumComputingDefects2010,
  title = {Quantum Computing with Defects},
  author = {Weber, J. R. and Koehl, W. F. and Varley, J. B. and Janotti, A. and Buckley, B. B. and {Van de Walle}, C. G. and Awschalom, D. D.},
  year = 2010,
  month = may,
  journal = {Proceedings of the National Academy of Sciences},
  volume = {107},
  number = {19},
  pages = {8513--8518},
  publisher = {Proceedings of the National Academy of Sciences},
  doi = {10.1073/pnas.1003052107}
}

@article{weilerSpinPumpingCoherent2012,
  title = {Spin {{Pumping}} with {{Coherent Elastic Waves}}},
  author = {Weiler, M. and Huebl, H. and Goerg, F. S. and Czeschka, F. D. and Gross, R. and Goennenwein, S. T. B.},
  year = 2012,
  month = apr,
  journal = {Physical Review Letters},
  volume = {108},
  number = {17},
  pages = {176601},
  doi = {10.1103/PhysRevLett.108.176601}
}

@article{wintzMagneticVortexCores2016,
  title = {Magnetic Vortex Cores as Tunable Spin-Wave Emitters},
  author = {Wintz, Sebastian and Tiberkevich, Vasil and Weigand, Markus and Raabe, J{\"o}rg and Lindner, J{\"u}rgen and Erbe, Artur and Slavin, Andrei and Fassbender, J{\"u}rgen},
  year = 2016,
  month = nov,
  journal = {Nature Nanotechnology},
  volume = {11},
  number = {11},
  pages = {948--953},
  issn = {1748-3395},
  doi = {10.1038/nnano.2016.117},
  copyright = {2016 Springer Nature Limited}
}

@article{wojewodaUnidirectionalPropagationZeromomentum2024,
  title = {Unidirectional Propagation of Zero-Momentum Magnons},
  author = {Wojewoda, Ond{\v r}ej and Holobr{\'a}dek, Jakub and Pavelka, Dominik and Pribytova, Ekaterina and Kr{\v c}ma, Jakub and Kl{\'i}ma, Jan and Panda, Jaganandha and Michali{\v c}ka, Jan and Lednick{\'y}, Tom{\'a}{\v s} and Chumak, Andrii V. and Urb{\'a}nek, Michal},
  year = 2024,
  month = sep,
  journal = {Applied Physics Letters},
  volume = {125},
  number = {13},
  pages = {132401},
  issn = {0003-6951},
  doi = {10.1063/5.0218478}
}

@article{wolskiDissipationBasedQuantumSensing2020,
  title = {Dissipation-{{Based Quantum Sensing}} of {{Magnons}} with a {{Superconducting Qubit}}},
  author = {Wolski, S. P. and {Lachance-Quirion}, D. and Tabuchi, Y. and Kono, S. and Noguchi, A. and Usami, K. and Nakamura, Y.},
  year = 2020,
  month = sep,
  journal = {Physical Review Letters},
  volume = {125},
  number = {11},
  pages = {117701},
  doi = {10.1103/PhysRevLett.125.117701}
}

@article{xiangIntracityQuantumCommunication2017,
  title = {Intracity {{Quantum Communication}} via {{Thermal Microwave Networks}}},
  author = {Xiang, Ze-Liang and Zhang, Mengzhen and Jiang, Liang and Rabl, Peter},
  year = 2017,
  month = mar,
  journal = {Physical Review X},
  volume = {7},
  number = {1},
  pages = {011035},
  publisher = {American Physical Society},
  doi = {10.1103/PhysRevX.7.011035}
}

@article{xueDirectionalEntanglementSpinorbit2025,
  title = {Directional Entanglement of Spin-Orbit Locked Nitrogen-Vacancy Centers by Magnons},
  author = {Xue, Zhiping and Zou, Ji and Cai, Chengyuan and Bauer, Gerrit E. W. and Yu, Tao},
  year = 2025,
  month = sep,
  journal = {Physical Review B},
  volume = {112},
  number = {9},
  pages = {094438},
  publisher = {American Physical Society},
  doi = {10.1103/9rvd-h7yq}
}

@article{xuQuantumControlSingle2023,
  title = {Quantum {{Control}} of a {{Single Magnon}} in a {{Macroscopic Spin System}}},
  author = {Xu, Da and Gu, Xu-Ke and Li, He-Kang and Weng, Yuan-Chao and Wang, Yi-Pu and Li, Jie and Wang, H. and Zhu, Shi-Yao and You, J. Q.},
  year = 2023,
  month = may,
  journal = {Physical Review Letters},
  volume = {130},
  number = {19},
  pages = {193603},
  doi = {10.1103/PhysRevLett.130.193603}
}

@misc{yamamotoTenSecondElectronSpinCoherence2026,
  title = {Ten-{{Second Electron-Spin Coherence}} in {{Isotopically Engineered Diamond}}},
  author = {Yamamoto, Takashi and {van Ommen}, H. Benjamin and Schymik, Kai-Niklas and {de Zoeten}, Beer and Onoda, Shinobu and Saiki, Seiichi and Ohshima, Takeshi and {Arjmandi-Tash}, Hadi and Vollmer, Ren{\'e} and Taminiau, Tim H.},
  year = 2026,
  month = apr,
  howpublished = {https://arxiv.org/abs/2604.07439v1},
  langid = {english}
}

@article{yinHybridOptomechanicalSystems2015,
  title = {Hybrid Opto-Mechanical Systems with Nitrogen-Vacancy Centers},
  author = {Yin, ZhangQi and Zhao, Nan and Li, TongCang},
  year = 2015,
  month = may,
  journal = {Science China Physics, Mechanics \& Astronomy},
  volume = {58},
  number = {5},
  pages = {1--12},
  issn = {1869-1927},
  doi = {10.1007/s11433-015-5651-1}
}

@article{yuanQuantumMagnonicsWhen2022,
  title = {Quantum Magnonics: {{When}} Magnon Spintronics Meets Quantum Information Science},
  shorttitle = {Quantum Magnonics},
  author = {Yuan, H. Y. and Cao, Yunshan and Kamra, Akashdeep and Duine, Rembert A. and Yan, Peng},
  year = 2022,
  month = jun,
  journal = {Physics Reports},
  series = {Quantum Magnonics: {{When}} Magnon Spintronics Meets Quantum Information Science},
  volume = {965},
  pages = {1--74},
  issn = {0370-1573},
  doi = {10.1016/j.physrep.2022.03.002}
}

@article{zakeriAsymmetricSpinWaveDispersion2010,
  title = {Asymmetric {{Spin-Wave Dispersion}} on {{Fe}}(110): {{Direct Evidence}} of the {{Dzyaloshinskii-Moriya Interaction}}},
  shorttitle = {Asymmetric {{Spin-Wave Dispersion}} on {{Fe}}(110)},
  author = {Zakeri, {\relax Kh}. and Zhang, Y. and Prokop, J. and Chuang, T.-H. and Sakr, N. and Tang, W. X. and Kirschner, J.},
  year = 2010,
  month = mar,
  journal = {Physical Review Letters},
  volume = {104},
  number = {13},
  pages = {137203},
  doi = {10.1103/PhysRevLett.104.137203}
}

@article{zarerameshtiCavityMagnonics2022,
  title = {Cavity Magnonics},
  author = {Zare Rameshti, Babak and Viola Kusminskiy, Silvia and Haigh, James A. and Usami, Koji and {Lachance-Quirion}, Dany and Nakamura, Yasunobu and Hu, Can-Ming and Tang, Hong X. and Bauer, Gerrit E. W. and Blanter, Yaroslav M.},
  year = 2022,
  month = sep,
  journal = {Physics Reports},
  series = {Cavity {{Magnonics}}},
  volume = {979},
  pages = {1--61},
  issn = {0370-1573},
  doi = {10.1016/j.physrep.2022.06.001}
}

@article{zhangCavityMagnomechanics2016,
  title = {Cavity Magnomechanics},
  author = {Zhang, Xufeng and Zou, Chang-Ling and Jiang, Liang and Tang, Hong X.},
  year = 2016,
  month = mar,
  journal = {Science Advances},
  volume = {2},
  number = {3},
  pages = {e1501286},
  doi = {10.1126/sciadv.1501286}
}

@article{zhangOptomagnonicWhisperingGallery2016,
  title = {Optomagnonic {{Whispering Gallery Microresonators}}},
  author = {Zhang, Xufeng and Zhu, Na and Zou, Chang-Ling and Tang, Hong X.},
  year = 2016,
  month = sep,
  journal = {Physical Review Letters},
  volume = {117},
  number = {12},
  pages = {123605},
  doi = {10.1103/PhysRevLett.117.123605}
}

@article{zhangStronglyCoupledMagnons2014,
  title = {Strongly {{Coupled Magnons}} and {{Cavity Microwave Photons}}},
  author = {Zhang, Xufeng and Zou, Chang-Ling and Jiang, Liang and Tang, Hong X.},
  year = 2014,
  month = oct,
  journal = {Physical Review Letters},
  volume = {113},
  number = {15},
  pages = {156401},
  doi = {10.1103/PhysRevLett.113.156401}
}
\end{document}